%% file: main3.tex
\documentclass[twocolumn,aps,prx,superscriptaddress,amsfonts,longbibliography]{revtex4-1}

\usepackage{physics,tensor,siunitx}
\usepackage{amsmath}
\usepackage{amssymb}
\usepackage{graphicx}
\usepackage{bm}
\usepackage{array}
\usepackage{dcolumn}
\usepackage{hepparticles}
\usepackage{comment}
\usepackage{heppennames}
\usepackage{hepnicenames}
\usepackage{epstopdf}
\usepackage{newfloat}
\usepackage{nicefrac}
\DeclareFloatingEnvironment[name={Extended Data Figure}]{suppfigure}

\usepackage{tikz}
\usetikzlibrary{decorations.pathmorphing} 
\usetikzlibrary{fit}					
\usetikzlibrary{backgrounds}	
\usetikzlibrary{shapes.geometric} %
\usetikzlibrary{positioning}
\usetikzlibrary{arrows.meta}

\begin{document}

\title{Quantum Computation of Hydrogen Bond Dynamics and Vibrational Spectra}

\author{Philip Richerme}
\affiliation{Indiana University Department of Physics, Bloomington, Indiana 47405, USA}
\affiliation{Indiana University Quantum Science and Engineering Center, Bloomington, Indiana 47405, USA}
\author{Melissa C. Revelle}
\affiliation{Sandia National Laboratories, Albuquerque, New Mexico 87123, USA}
\author{Debadrita Saha}
\affiliation{Indiana University Department of Chemistry, Bloomington, Indiana 47405, USA}
\author{Miguel Angel Lopez-Ruiz}
\affiliation{Indiana University Department of Chemistry, Bloomington, Indiana 47405, USA}
\author{Anurag Dwivedi}
\affiliation{Indiana University Department of Chemistry, Bloomington, Indiana 47405, USA}
\author{Sam A. Norrell}
\affiliation{Indiana University Department of Physics, Bloomington, Indiana 47405, USA}
\author{Christopher G. Yale}
\affiliation{Sandia National Laboratories, Albuquerque, New Mexico 87123, USA}
\author{Daniel Lobser}
\affiliation{Sandia National Laboratories, Albuquerque, New Mexico 87123, USA}
\author{Ashlyn D. Burch}
\affiliation{Sandia National Laboratories, Albuquerque, New Mexico 87123, USA}
\author{Susan M. Clark}
\affiliation{Sandia National Laboratories, Albuquerque, New Mexico 87123, USA}
\author{Jeremy M. Smith}
\affiliation{Indiana University Department of Chemistry, Bloomington, Indiana 47405, USA}
\author{Amr Sabry}
\affiliation{Indiana University Quantum Science and Engineering Center, Bloomington, Indiana 47405, USA}
\affiliation{Indiana University Department of Computer Science, Bloomington, Indiana 47405, USA}
\author{Srinivasan S. Iyengar}
\affiliation{Indiana University Quantum Science and Engineering Center, Bloomington, Indiana 47405, USA}
\affiliation{Indiana University Department of Chemistry, Bloomington, Indiana 47405, USA}
\date{\today}

\begin{abstract}
Calculating the observable properties of chemical systems is often classically intractable and is widely viewed as a promising application of quantum information processing. Yet one of the most common and important chemical systems in nature – the hydrogen bond – has remained a challenge to study using quantum hardware on account of its anharmonic potential energy landscape. Here, we introduce a framework for solving hydrogen-bond systems and more generic chemical dynamics problems using quantum logic. We experimentally demonstrate a proof-of-principle instance of our method using the QSCOUT ion-trap quantum computer, in which we experimentally drive the ion-trap system to emulate the quantum wavepacket of the shared-proton within a hydrogen bond. Following the experimental creation of the shared-proton wavepacket, we then extract measurement observables such as its time-dependent spatial projection and its characteristic vibrational frequencies to spectroscopic accuracy ($3.3$ cm$^{-1}$ wavenumbers, corresponding to $> 99.9\%$ fidelity). Our approach introduces a new paradigm for studying the quantum chemical dynamics and vibrational spectra of molecules, and when combined with existing algorithms for electronic structure, opens the possibility to describe the complete behavior of complex molecular systems with unprecedented accuracy.

\end{abstract}
\maketitle

\section{Introduction}
  Hydrogen bonds and hydrogen transfer dynamics are fundamental to most biological, materials and atmospheric systems \cite{htrans,Breslow-Organic-Water,Marcus-Onwater,Yano-Yachandra, Lubitz,Sshb-Serine-Protease}, including the anomalous Grotthuss-like behavior in  water \cite{grotthuss,Johnson-Jordan-21mer} and fuel cells \cite{voth-nafion}, proton transfer in photosynthesis \cite{Meyer-2005_IC_44_6802,Yachandra-Yano-Water-splitting}, and   nitrogen-fixation \cite{Hoffman-N2-H2-redelim}.  
  Hydrogen bonds are characterized by the interactions of a single hydrogen atom confined to a ``box'' created by electronegative donor and acceptor atoms that regulate the extent to which the bond is polarized. Since the shape of this box is inherently anharmonic \cite{HDMeyer-Zundel-1,qwaimd-wavelet,Johnson-2007-Science-Me2OH+,Me2O2D+Xiaohu,kamarchik2009coupled}, the commonly-used approach of modeling chemical bonds as harmonic springs \cite{Pople-2nd-derivs,Pople-2nd-derivs-freqs} is known to drastically fail for hydrogen-bonded systems \cite{HDMeyer-Zundel-1,qwaimd-wavelet,Johnson-Jordan-Zundel-Science,Johnson-Jordan-21mer,Me2O2D+Xiaohu}. As a result, calculating the full quantum wavepacket dynamics \cite{HDMeyer-Zundel-1,qwaimd-wavelet} is often the only approach to accurately predict the chemical behavior of such complex systems. Unfortunately multi-dimensional quantum nuclear dynamics problems require exponential computational resources to achieve predictive results, forcing existing studies to sacrifice accuracy for efficiency.

Solving proton- or hydrogen-transfer dynamics problems using quantum-inspired approaches has also posed a challenge to date. While promising quantum algorithms \cite{abrams1997simulation,aspuru2005simulated,wang2008quantum,whitfield2011simulation,wecker2015progress,mcclean2016theory,o2016scalable,parrish2019quantum,frag-QC-Harry,smart2021quantum} and experiments \cite{lanyon2010towards,lu2011simulation,peruzzo2014variational,kandala2017hardware,hempel2018quantum,nam2020ground,google2020hartree} have addressed systems with strongly-correlated electrons, these techniques are incapable of describing quantum nuclear effects within molecules. Likewise, while pioneering work has performed quantum simulations of vibronic spectra \cite{sparrow2018simulating,wang2020efficient,huh2015boson} or wavepacket evolution through conical intersections \cite{wang2022observation}, these studies have been limited to molecular potentials approximated as harmonic (inapplicable to hydrogen-bonded systems) evaluated at a single molecular geometry.

Here, we introduce a suite of techniques for performing accurate quantum nuclear dynamics calculations on a single Born-Oppenheimer potential energy surface using quantum hardware. Our approach exploits the fact that quantum nuclear dynamics problems may contain far fewer than $\mathcal{O}(4^n)$ independent elements when represented as a $2^n \times 2^n$ matrix, requiring a dramatically reduced number of operations when mapped to quantum simulation hardware \cite{saha2021mapping}. While in this work we focus on mapping simple one-dimensional hydrogen-bond systems, we also provide methods which extend this framework to address more general chemical systems which may include multiple correlated quantum nuclear dimensions.

To provide a proof-of-principle demonstration of our approach, we emulate the time dynamics of a shared-proton wavepacket evolving within a hydrogen-bonded system \cite{Sshb-Enzymes} using a two-qubit quantum computer. For specificity we choose an exemplar molecule, monoprotonated bis(1,8-dimethylamino)naphthalene (DMANH$^+$), which has a shared proton bound by a smooth double-well potential spanning the donor-acceptor range and exhibits a structure common to most strong hydrogen-bonded systems. Our experiment implements the effective nuclear Hamiltonian of this hydrogen bonded system by mapping it to a trapped-ion quantum computer with site-specific preparation, control, and readout \cite{clark2021engineering}. We apply discrete quantum logic gates to drive the system's unitary time-evolution, from which we directly determine the spatial and temporal projection of the shared-proton wavepacket, its characteristic vibrational (anharmonic) frequencies, and the complete energy eigenspectrum of the applied nuclear Hamiltonian. Our experiment shows exceptional agreement with the corresponding classical calculations of the shared-proton vibrational and energy spectra for this model problem, with a $0.1\%$ uncertainty ($\approx$ 3 cm$^{-1}$ error in the resultant vibrational spectrum). The experimental techniques described here are directly scalable to larger numbers of qubits, and following the framework we introduce below, may be applied to study more complex chemical problems with multiple nuclear degrees of freedom.

The paper is organized as follows. In Section \ref{theory} we review the framework for mapping chemical dynamics problems to quantum hardware. Section \ref{dynamics} presents our ion-trap based emulation of wavepacket dynamics, corresponding to the shared-proton motion of a single hydrogen-bonding mode in DMANH$^+$. The resulting experimentally-determined vibrational spectrum is shown and compared with the results of a classical calculation in Section \ref{spectrum}. In Section \ref{tensor}, we show how our approach may be generalized and extended to more complex chemical systems with arbitrary number of vibrational degrees of freedom. We summarize with concluding remarks in Section \ref{conclusions}.

\section{Theoretical Framework for Mapping Chemical Dynamics}
\label{theory}
The full quantum treatment of chemical dynamics problems requires solving the Schr\"odinger propagation of nuclear wavepackets evolving under the system Hamiltonian. When expressed on classical hardware, this is believed to be an exponentially-scaling problem with system size \cite{hibbs,feit1,MCTDH-Meyer1,Feynman-Comp}. The general problem of chemical dynamics is complicated by two primary challenges: (a) The classical computational cost associated with obtaining accurate electronic potential energy surfaces may grow exponentially with the number of nuclear degrees of freedom \cite{Bowman-IRPC-PES,qwaimd-wavelet,DieterPOTFIT}; 
(b) Both the classical storage of the quantum unitary propagator and quantum wavepackets, as well as the time-evolution of the propagator acting on the quantum states, may also require exponential resources with system size \cite{BastiaTT-SOFT,tensor_network,boris2007tensorProduct,MLMCTDHWang,Nicole-TN,frag-TN-Anup}.

In this section, we present a framework which directly addresses challenge (b), by expressing generic nuclear Hamiltonians in a form amenable to implementation on quantum hardware. Specifically, we show how the nuclear dynamics on a single Born-Oppenheimer potential energy surface may be mapped to an interacting configuration of quantum bits, as may be found in spin-lattice quantum simulators or universal quantum computers. This framework is scalable to the continuum limit along a single nuclear dimension and provides a foundation for treating multiple correlated nuclear degrees of freedom (see Section \ref{tensor}). We note that in parallel, aspects of challenge (a) are being addressed by several groups seeking to implement electronic-structure calculations on quantum hardware \cite{lanyon2010towards,lu2011simulation,peruzzo2014variational,kandala2017hardware,hempel2018quantum,nam2020ground,google2020hartree,frag-QC-Harry} and by groups seeking to reduce the potentially exponential scaling of electronic structure calculations needed for accurate potential energy surfaces \cite{frag-TN-Anup}. 

\subsection{Structure of the Nuclear Hamiltonian}
For both classical- and quantum-computational approaches to studying chemical dynamics problems, the initial step is to express the nuclear Hamiltonian in some suitable basis, such as a discrete coordinate basis (Fig. \ref{f1}a) that is commonly used in quantum dynamics studies \cite{feit1,youhong,stuart-Nature,Leforestier-Lanczos,talezer,SKGray-RealWP,SincDVR}. In our study, we interpolate a lattice of \mbox{$2^N$} equally-spaced points between the donor/acceptor groups, then choose a set of $2^N$ basis functions corresponding to Dirac delta-functions centered on those lattice sites. The shape of the shared-proton wavepacket at any time is thus represented by a ket vector, with its time dependence governed by unitary Schr\"odinger evolution under the system Hamiltonian. 

In this discrete basis, the diagonal elements of the shared-proton Hamiltonian encode an effective double-well potential energy surface arising from the surrounding nuclear frameworks and electronic structure. This potential energy surface, local in the coordinate representation, includes explicit electron correlation and is treated using density functional theory. For a specific molecular system, the calculated shape of the double-well potential depends upon the level of electronic structure theory used to depict electron correlation, as well as the chosen grid spacing used to represent the nuclear degrees of freedom.

The off-diagonal elements of the shared-proton Hamiltonian encode the nuclear kinetic energy and are determined here using distributed approximating functionals (DAFs) \cite{DAFprop-PRL,discreteDAF,saha2021mapping,qwaimd-TCAreview}:
\begin{align}
K(x,x^{\prime}) =& K(\left\vert x-x^{\prime}\right\vert) =
\frac{-\hbar^2}{4m\sigma^3\sqrt{2\pi}}
\exp \left\{ -\frac{ {\left( x - x^\prime \right)}^2}
{2 {\sigma}^2} \right\} \nonumber \\ & \sum_{n=0}^{M_{DAF}/2} {\left( \frac{-1}{4} \right)}^n \frac{1}{n!} 
H_{2n+2} \left( \frac{ x - x^\prime }{ \sqrt{2} \sigma} \right).
\label{DAFfreeprop+derivative}
\end{align}
 The quantities $\left\{ H_{2n+2}\left( \frac{ x - x^\prime }{ \sqrt{2} \sigma} \right) \right\}$ are the even order Hermite polynomials that only depend on the spread separating the grid basis vectors, $\ket{x}$ and $\ket{x^{\prime}}$, and $M_{DAF}$ and $\sigma$ are parameters that together determine the accuracy and efficiency of the resultant approximate kinetic energy operator. Appendix D in Ref. \onlinecite{saha2021mapping} provides a brief summary of the DAF approach for approximating functions, their derivatives and propagated forms. 

\subsection{Mapping from Molecule to Quantum Hardware}
Together, the potential and kinetic energy constitute the molecular Hamiltonian, written in the coordinate representation as
\begin{align}
    \bra{x} {{\cal H}}^{Mol} \ket{x^{\prime}} 
    = 
    K(x,x^{\prime}) + V(x)\delta \left( {x-x^{\prime}} \right)
    \label{HDAF-molQD}
\end{align}
Here we focus on one-dimensional implementations of Eq. \ref{HDAF-molQD}, with multi-dimensional generalizations presented in Section \ref{tensor}. Underlying symmetries in the nuclear Hamiltonian allow it to be written in a form more amenable for efficient experimental realizations \cite{Debadrita-Mapping-1D-3Qubits}. The symmetric double-well potential seen by the shared-proton wavepacket, for instance, leads to a reflection symmetry in the diagonal elements of the Hamiltonian. In addition, the kinetic energy operator of the shared-proton wavepacket, when represented using DAFs, imprints a Toeplitz structure to the off-diagonal elements for which each sub-diagonal has a constant value. Matrices with these two key properties can be transformed into a block-diagonal form by applying a Givens rotation operation to the computational basis \cite{Debadrita-Mapping-1D-3Qubits,sm}. The diagonal blocks of the transformed Hamiltonian $\tilde{\mathcal{H}}$ may then be written
   \begin{align}
      \tilde{\mathcal{H}}^{Mol}_{il} &= \left[ {K}(x_{i},x_{l}) + \alpha_{i} {K}(x_{i},x_{n-l}) \right] + \nonumber \\ &  \hphantom{= \frac{1}{2}(}\frac{1}{2} \left[ V(x_{i}) + V(x_{n-l}) \right] \delta_{i,l}
      \label{Htilde-ii}
  \end{align}
  with off-diagonal blocks given by 
  \begin{align}
      \tilde{\mathcal{H}}^{Mol}_{il} =& 
      \frac{1}{2} \left[ V(x_{i}) - V(x_{n-l}) \right]  
      \delta_{i,n-l}
      \label{Htilde-il-od}
  \end{align}
  where the grid point indices take values from 0 to $n= 2^N-1$, and $N$ is the number of qubits used to encode the system. The quantity $\alpha_{i}=\text{sgn}\left[i-(n/2)\right]$. For symmetric potentials, as are common in many hydrogen-bonded systems, the off-diagonal blocks of the transformed Hamiltonian (Eq. \ref{Htilde-il-od}) are identically zero which yields a more efficient representation of the original Hamiltonian.
  
The net effect of this mapping is to recast the original $2^N \times 2^N$ nuclear Hamiltonian as one with two sub-blocks of dimension $2^{N-1} \times 2^{N-1}$, with each sub-block operating independently on half of the Hilbert space. Within each sub-space, the transformed Hamiltonian blocks operate on a transformed grid basis given by
  \begin{align}
        \ket{\tilde{x}_{i}} &\equiv \frac{1}{\sqrt{2}}\left[{\ket{x_{i}} + \ket{x_{n-i}}}\right], \quad 0 \leq i < (n+1)/2 \label{Givens-transformed-basis-1} \\
     &\equiv \frac{1}{\sqrt{2}}\left[{\ket{x_{i}} - \ket{x_{n-i}}}\right], \quad (n+1)/2 \leq i \leq n
     \label{Givens-transformed-basis}
\end{align}
Thus, the time-evolution of the full Hamiltonian can be implemented by performing time-evolution of both sub-blocks independently, then applying an inverse mapping to transform back to the original basis \cite{sm}.

\subsection{Implementing Nuclear Hamiltonians on Quantum Hardware}
Following the generation of the block-diagonal molecular Hamiltonian, further processing must be done to implement the time-evolution operator ($U(t)=e^{-iHt/\hbar}$) on an initialized register of trapped-ion qubits. Both sub-blocks of the block-diagonal Hamiltonian can be quantum-simulated independently, since the sub-blocks act in independent Hilbert spaces. Below, we discuss two approaches for implementing this unitary propagator on a collection of interacting quantum bits.

In principle, for one-dimensional nuclear dynamics problems, an infinite number of qubits will be required to reach the `continuum limit' for which the spacing between discrete gridpoints approaches zero and perfect accuracy is achieved. Yet, for each nuclear dimension, only 6-7 qubits will be required in practice to reach the level of precision required for the accurate determination of chemical dynamics and vibrational spectra. This is motivated by comparing the average de Broglie wavelength of light nuclei (generally fractions of Angstroms) to the spatial extent of an average chemical bond (generally 1-2 Angstroms). Hence, a grid discretization distance that is ~100 times smaller than the length of the bond is sufficient to describe the relevant dynamics to beyond chemical or spectroscopic accuracy \cite{youhong,SKGray-RealWP,Bowman-IRPC-PES,qwaimd-wavelet}. Using 7 qubits to interpolate the single dimension with $2^7 = 128$ gridpoints easily satisfies this criterion; when combined with extensions to multiple dimensions (section \ref{tensor}), chemical systems with multiple correlated degrees of freedom may be addressed as well.

\subsubsection{Ising-Model Implementations}
\label{isingimp}
The quantum Ising Model is a foundational paradigm for describing the behavior of interacting spin systems, and it has been studied widely on diverse quantum hardware platforms such as trapped ions \cite{monroe2021programmable}, Rydberg atoms \cite{labuhn2016tunable}, polar molecules \cite{yan2013observation}, cold atomic gases \cite{bloch2012quantum}, and superconducting circuits \cite{barends2016digitized}. In its full implementation, the quantum Ising Model Hamiltonian with local magnetic fields may be written:
\begin{equation}
\label{eq:tfim}
H=\sum_{\gamma}\sum_{i<j} J_{ij}^\gamma \sigma_i^\gamma \sigma_j^\gamma + \sum_{\gamma}\sum_{i}B_i^\gamma\sigma_i^\gamma
\end{equation}
where $\gamma \in {(x,y,z)}$, $J_{ij}^\gamma$ is the coupling between sites $i$ and $j$ along the $\gamma$ direction, $B_i^\gamma$ is the local magnetic field at site $i$ along the the $\gamma$ direction, and the quantities $\left\{ \sigma_i^\gamma \right\}$ are the Pauli spin operators acting on the $i^{th}$ lattice site along the $\gamma$-direction of the Bloch sphere. 

When the basis states of the Ising Model are properly permuted \cite{Debadrita-Mapping-1D-3Qubits,sm}, the Hamiltonian in Eq. \ref{eq:tfim} yields a block-diagonal structure identical to that of the transformed molecular Hamiltonian $\tilde{\mathcal{H}}^{Mol}$. This shared Hamiltonian structure enables the molecular Hamiltonian to be emulated on Ising spin-lattice simulators following specific choices for the $J_{ij}^\gamma$ interactions and $B_i^\gamma$ local fields (the full procedure is described in \cite{Debadrita-Mapping-1D-3Qubits,sm}).

For most quantum hardware platforms, the most straightforward implementation of the desired $J_{ij}^\gamma$ interactions and $B_i^\gamma$ fields will be through a digital quantum simulation approach \cite{lloyd1996universal}. In this approach each individual coupling or local field is applied stroboscopically, and $n$ multiple cycles are repeated to evolve the system dynamics. The error arising from such `Trotterization' scales as $\mathcal{O}(T^2)$ \cite{lloyd1996universal} and may be kept arbitrarily small by increasing $n$ proportionally. Since there are of order $\mathcal{O}(N^2)$ Ising-type couplings between $N$ qubits, the total computational cost remains of order $\mathcal{O}(N^2)$ in the system size for fixed error.

\subsubsection{Circuit-Model Implementations}
Given the transformed block-diagonal molecular Hamiltonian $\tilde{\mathcal{H}}^{Mol}$, each sub-block may be directly decomposed into a sequence of single- and two-qubit gates for implementation on a universal quantum computer.
We emphasize that the decomposition of general unitary matrices scales poorly, requiring $\mathcal{O}(4^N)$ quantum operations to decompose a sub-block of $N$ qubits. Nevertheless, for small numbers of qubits (as in Section \ref{dynamics}), this approach requires fewer quantum operations compared to the `Trotterization' approach described above. For larger numbers of qubits, the structure and sparsity of the molecular Hamiltonian in Eqs. (\ref{DAFfreeprop+derivative}) and (\ref{HDAF-molQD}) may enable more efficient circuit decompositions compared to the general case; when extended to multiple degrees of freedom as discussed in Section \ref{tensor}, this may lead to a near-linear scaling of complexity as the number of quantum nuclear dimensions is increased. This potential quantum advantage will be explored and probed experimentally in future publications.

\begin{figure*}[t!]
\includegraphics[width=15.5cm]{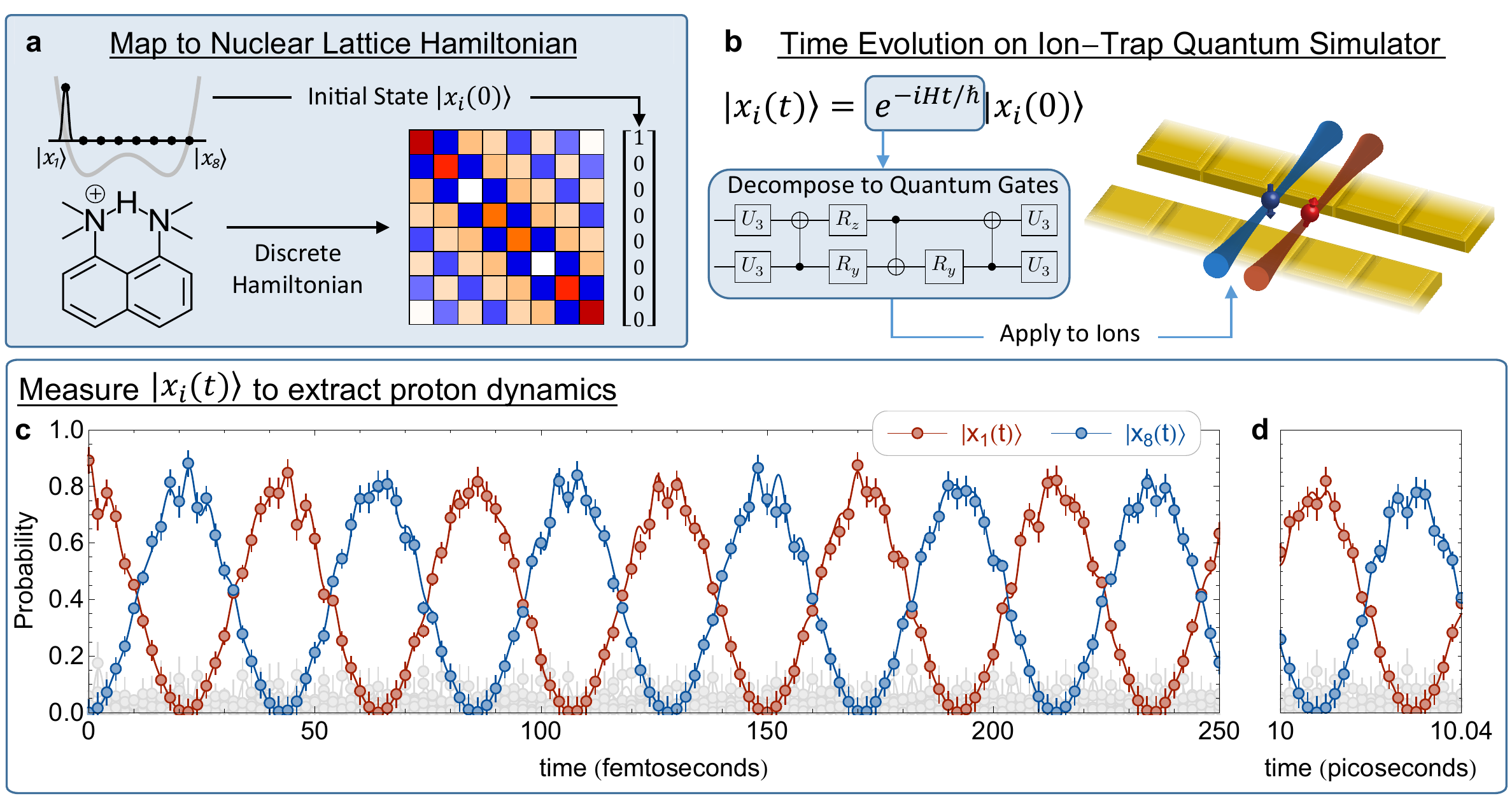}
\caption{\label{f1} Coherent dynamics of a shared-proton wavepacket. (a) The quantum dynamics of a short-strong hydrogen bonded system are mapped onto a discrete basis. Each basis state represents a shared-proton wavepacket occupying one of the equally-spaced lattice sites between the donor/acceptor frameworks. The discrete Hamiltonian is formed from writing the kinetic and potential energy operators for the shared-proton in this system. (b) Time evolution under the nuclear Hamiltonian is achieved by applying a unitary propagator to an initial wavepacket state. The unitary operator is decomposed into single- and two-qubit quantum gates which are applied to a trapped-ion quantum computer. (c) The quantum dynamics of the trapped-ion system emulate those of the shared proton wavepacket. When a wavepacket is initialized near the left side of the anharmonic potential well, its dynamics are found to oscillate coherently between the donor and acceptor groups. Solid lines show the exact numerical solution to the Schr\"odinger equation including decoherence effects. (d) Our trapped-ion experiment remains coherent for many hundreds of emulated shared-proton oscillations with no visible reduction in contrast.}
\end{figure*}

For the 2-qubit experimental demonstration described in Section \ref{dynamics}, we utilize the Cartan Decomposition to optimally decompose a general $4\times 4$ unitary matrix sub-block into a sequence of seven single-qubit rotations and three controlled-NOT (CNOT) gates \cite{vatan2004optimal,tucci2005introduction}. A schematic of this quantum circuit is shown in Fig. \ref{f1}b. Controlled-NOT gates are realized via a decomposition into single qubit rotations sandwiching a single XX-type two-qubit M\o lmer-S\o rensen interaction \cite{maslov2017basic,molmer1999multiparticle} so that they may be applied to trapped-ion qubits. Local unitary operations, $U=e^{i\alpha}R_\phi(\theta)$, are further decomposed into standard quantum logic rotation gates through a general transformation $U=e^{i\alpha}R_z(\beta)R_y(\gamma)R_z(\delta)$ \cite{nielsen-and-chuang}. The Euler rotation angles $\beta$, $\gamma$, and $\delta$ are evaluated \textit{modulo} $2\pi$ and constrained on the interval $[-\pi,\pi]$ before being passed to the quantum computer. This minimizes gate timing errors and geometrically corresponds to performing the minimal equivalent rotation for each single-qubit operation. 

The above methods factorize a general $4\times 4$ unitary matrix sub-block into a series of single-qubit rotations and three M\o lmer-S\o rensen interactions. The parameters characterizing these quantum logic gates define the time-evolution operator $U(t)$ for a given evolution time $t$ and are calculated explicitly for each Hamiltonian evolution timestep. Since the structure of the quantum circuit remains fixed for all timesteps, differing only in the single-qubit rotation angles, the overall quantum circuit fidelity is approximately constant for any $U(t)$ with errors arising primarily from the three CNOT operations.



\section{Wavepacket Dynamics on Ion-Trap Systems}
\label{dynamics}

To emulate the chemical dynamics of this hydrogen-bonded system, lasers are used to prepare and manipulate the internal electronic states of trapped atomic ions (Fig. \ref{f1}b). Qubit levels are encoded in the $^2$S$_{1/2}\ket{F=0,m_F=0}$ and $\ket{F=1,m_F=0}$ hyperfine `clock' states of $^{171}$Yb$^+$ ions, denoted $\ket{0}$ and $\ket{1}$, respectively \cite{olmschenk2007manipulation}. State preparation proceeds by optically pumping all qubits to $\ket{0}$, then performing local rotations to generate a state corresponding to an initial shared-proton wavepacket. To find the shared-proton wavepacket at a later time $t$, we time-evolve the initial state under the unitary propagator $U(t)=e^{-iHt/\hbar}$, where $H$ is the ion-trap Hamiltonian that encodes the effective nuclear dynamics. While $H$ is generically expressed in terms of Ising-type XX and YY interactions \cite{saha2021mapping}, here we exploit the small system size to optimally decompose the propagator \mbox{$U(t)=e^{-iHt/\hbar}$}  into a sequence of seven single-qubit rotations (with time-dependent angles) and three controlled-NOT (CNOT) gates \cite{vatan2004optimal,tucci2005introduction,maslov2017basic,molmer1999multiparticle}. 

We generate single- and two-qubit gates by applying spin-dependent optical dipole forces to ions confined in a surface-electrode trap (Sandia HOA-2.1 \cite{maunz2016high}). Each ion is addressed by two laser beams near 355 nm: one tightly-focused individual-addressing beam and one global beam that targets all ions simultaneously. The beams are arranged such that their wavevector difference $\Delta \vec{k}$ lies along the transverse principal axes in plane with the trap surface. The two beams contain frequency components whose difference can be matched to the resonant transition of the ion qubit at $\nu_0 = 12.642819$ GHz to drive single-qubit gates, or detuned symmetrically from $\nu_0$ to drive M\o lmer-S\o rensen two-qubit interactions. Gate times for single-qubit $\pi/2$ flips are typically 10~$\mu$s with $99.5\%$ fidelity, while typical M\o lmer-S\o rensen gates require 200~$\mu$s for full entanglement and a typical $97\%$ fidelity.

After initializing the ion qubits and applying gates to simulate Hamiltonian evolution for time $t$, the qubit states are measured to determine the time-evolved shape of the shared-proton wavepacket. At each timestep, measurements of the trapped-ion qubit states determine the probability of finding the shared-proton on each of the $2^N$ lattice sites. Detection of the final ion states is accomplished by capturing their spin-dependent fluorescence into a fiber array, where each fiber is coupled to an individual photomultiplier-tube (PMT). Site-resolved detection allows for discrimination of each qubit's logical $\ket{0}$ or $\ket{1}$ states with $99.0\%$ detection fidelity as well as the probability overlap with all possible basis states. Experiments are repeated 1000 times at each timestep to limit the statistical error contributions from quantum projection noise. The time-dependent proton dynamics are then determined by mapping the observed ion dynamics back to the discrete proton basis \cite{sm}.

In Figure \ref{f1}c, we highlight the probabilities of finding the shared-proton on the left-most ($\ket{x_1}$) and right-most ($\ket{x_8}$) lattice sites on the potential energy surface, when it is initialized in the $\ket{x_1}$ position. For this initial state, the shared-proton wavepacket is found to exhibit large-amplitude, coherent oscillations between the donor/acceptor groups as well as smaller-amplitude, higher-frequency oscillations; intermediate lattice sites contribute negligibly to the dynamics in this case (grey points in Fig. \ref{f1}c). We emphasize that the solid lines in Fig. \ref{f1}c are not fits to the data; rather, they are an exact numerical solution to the Schr\"odinger equation in the presence of known quantum gate infidelities.

In our experiments, the quantum circuit in Fig. \ref{f1}b is implemented with a $\sim90\%$ fidelity for any simulated evolution time $t$. This results in effective shared-proton oscillations that remain coherent for arbitrarily long times with no apparent decrease in contrast. Figure \ref{f1}d shows a continuation of the dynamics from Fig. \ref{f1}c, hundreds of oscillation periods after the wavepacket is initialized. 
Remarkably, since the gate errors are the same for each timestep, they do not affect the frequency information encoded in the time dynamics. For instance, we show below that the shared-proton vibrational frequencies may be determined to $>99.9\%$ accuracy despite the $\sim90\%$ overall circuit fidelity.

\begin{figure}[t]
\includegraphics[width=\columnwidth]{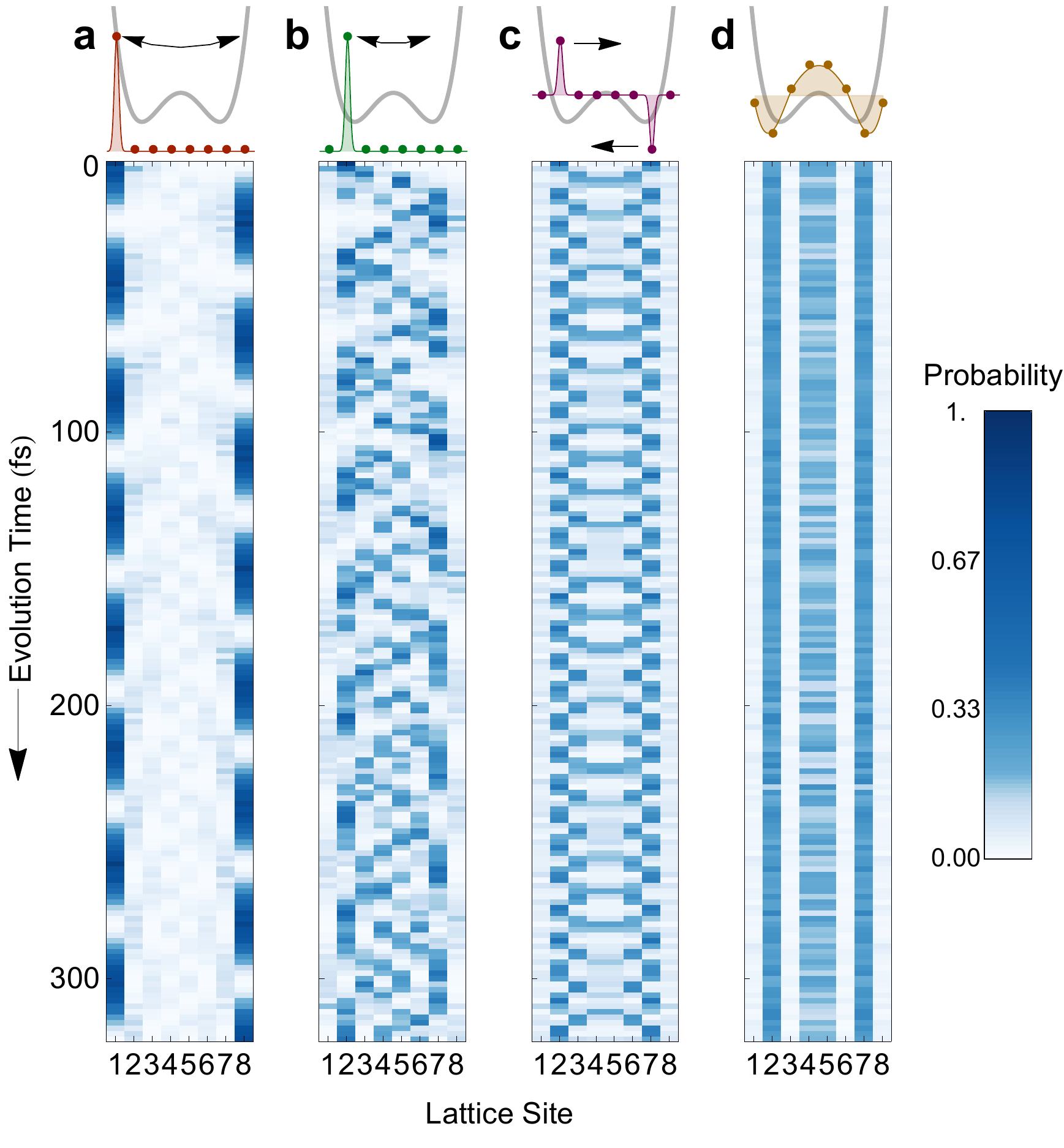}
\caption{\label{f2} Experimentally-determined dynamics of a shared-proton in DMANH$^+$. The observed time evolution depends on whether the shared-proton wavepacket is initialized (a) near the donor site; (b) near one of the potential energy surface minima; (c) split between the two minima with opposite phase; or (d) in an eigenstate of the double-well potential. Oscillations in both space and time are observed in all cases, except in (d), where the system remains in its eigenstate.}
\end{figure}

Initializing our system in different quantum states, which corresponds to initializing the shared-proton wavepacket on different lattice sites, leads to a variety of different time-dependent oscillations. In Figure \ref{f2}a, the effective shared-proton state is initialized to occupy the left-most lattice site at $t=0$. Just as in Fig. \ref{f1}c-d, the time dynamics reveal wavepacket jumps between the $\ket{x_1}$ and $\ket{x_8}$ lattice sites at a rate of $\approx$24 THz \mbox{(800 cm$^{-1}$)}. In contrast, when the shared-proton state is initialized on the second lattice site (Fig. \ref{f2}b), a continuous oscillation between sites $\ket{x_2}$ through $\ket{x_7}$ is observed with nearly-zero probability of occupying the extrema ($\ket{x_1}$ and $\ket{x_8}$). When positive and negative wavepacket components are introduced in the left- and right- wells (respectively), as in Fig. \ref{f2}c, strong oscillations are observed along with destructive interference effects near the center of the double well potential. Finally, when the initial proton wavepacket is prepared in an eigenstate of the nuclear Hamiltonian (Fig. \ref{f2}d), no time-dependent dynamics are expected nor observed.

\begin{figure*}[t]
\includegraphics[width=\textwidth]{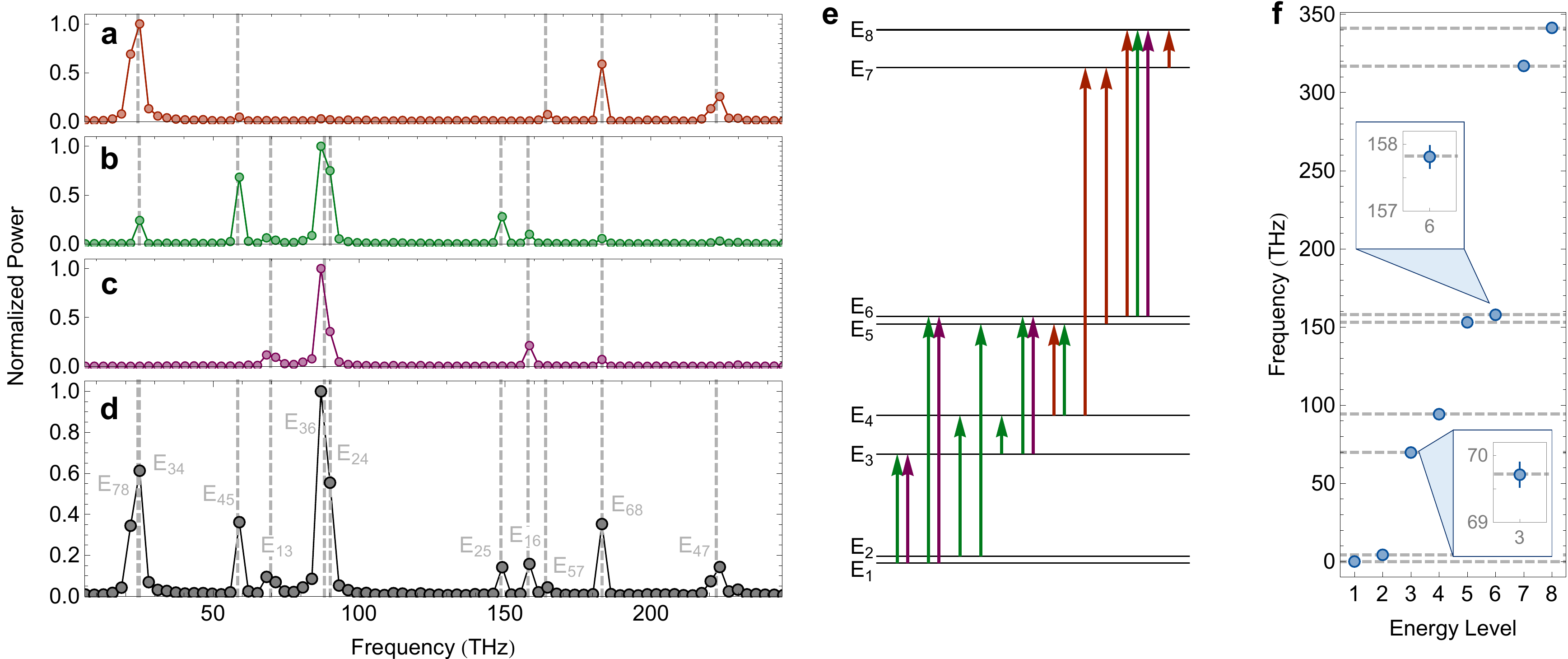}
\caption{\label{f3} Experimentally-determined frequency and energy spectra of a shared-proton in DMANH$^+$. (a)-(c) The time evolution data for the initial states in Fig. \ref{f2}a-c are Fourier-transformed to reveal frequency spectra of the shared-proton oscillation. Each peak corresponds to a frequency splitting between eigenstates of the discrete nuclear Hamiltonian. (d) The data in a-c are summed to produce a final frequency spectrum. Dashed grey lines and labels show predicted frequencies from exact diagonalization of the nuclear Hamiltonian. (e) The extracted frequencies from a-d, color-coded by their parent spectrum, allow the relative energies of all eigenstates to be experimentally determined. (f) The quantum-computed energy eigenstates of the nuclear Hamiltonian (blue dots) are compared to the exact-diagonalization result (dashed grey lines). Typical error bars (1 s.d.) are shown in the insets.}
\end{figure*}

\section{Determination of Vibrational Spectra from Wavepacket Dynamics}
\label{spectrum}
Our emulation of the shared-proton wavepacket dynamics enables a high-accuracy determination of its vibrational frequencies. Consider the nuclear Hamiltonian $H$, which has eigenstates $\phi_i(x)$ and energy eigenvalues $E_i$. Any chosen initial state $\chi(x,0)$ may be written in terms of this eigenstate basis, with corresponding time evolution:
\begin{equation}
    \chi(x,t)=\sum_i c_i(0) e^{-i E_i t/\hbar}\phi_i(x)
\end{equation}
Hence, the probability of finding the shared proton at position $x$ at time $t$ is given by
\begin{equation}
    |\chi(x,t)|^2 = \sum_{i,j} c_j^*(0) c_i(0) e^{i (E_j-E_i)t/\hbar} \phi_j^*(x) \phi_i(x)
\end{equation}
The time evolution of any initial state therefore is comprised of oscillations at all possible frequency differences between all pairs of energy eigenstates. For each frequency component, the strength is governed by the product of overlap amplitudes $c_j^*(0)c_i(0)$, that is, the coefficients comprising the initial wavepacket.

To extract the oscillation frequencies of the shared-proton wavepacket in our hydrogen-bonded system, we perform a Fourier transform of the measured time-evolution data presented in Figure \ref{f2}. Mathematically, this operation is equivalent to taking
\begin{equation}
\begin{split}
    \int e^{i\omega t} & |\chi(x,t)|^2 dt = \\
    &\sum_{i,j} \left[ \int e^{i \omega t} e^{i (E_j-E_i)t/\hbar} dt \right] c_j^* c_i  \phi_j^*(x) \phi_i(x) \\
    = &\sum_{i,j} \delta(\omega-(E_j-E_i)/\hbar) c_j^* c_i  \phi_j^*(x) \phi_i(x) \\
    \end{split}
    \label{FTequation}
\end{equation}
which produces peaks in the Fourier spectrum corresponding to frequency differences between energy eigenstates, $(E_j-E_i)/\hbar$. This expression is also related to the Fourier transform of the density matrix auto-correlation function, ${\text{Tr}}[\rho(0)\rho(t)]$ \cite{sm}. 

Fourier-transforms of the shared-proton time dynamics (Fig. \ref{f2}a-c) are shown in Fig. \ref{f3}a-c, along with a cumulative spectrum in Fig. \ref{f3}d. The extracted peaks provide a direct measurement of the energy eigenstate differences $E_{ij} \equiv (E_j-E_i)$, which show excellent agreement with the frequencies predicted from exact diagonalization of the nuclear Hamiltonian (grey dashed lines in Fig. \ref{f3}a-d). Note that since the initial wavepacket state in Figure \ref{f2}d was prepared in an eigenstate and contains minimal time dynamics, its Fourier transform contains no meaningful frequency information.

Using the measured energy differences $E_{ij}$ in Fig. \ref{f3}a-d, we experimentally reconstruct the full energy eigenspectrum of our nuclear Hamiltonian. Since the shared-proton basis has been discretized onto 8 lattice sites, our Hamiltonian contains 8 energy eigenvalues which are calculable and drawn to scale in Fig. \ref{f3}e. The set of measurements $\{E_{ij}\}$ from Fig. \ref{f3}a-d provides more information than necessary to determine the relative spacings of all energy levels. This overcompleteness arises from our multiple different wavepacket initializations, and it allows for multiple independent measurements of specific energy splittings and reduced error in our final results.

The vibrational energies obtained directly from the ions' time dynamics are compared in Fig. \ref{f3}f to the exact diagonalization results (dashed grey lines). Exceptional agreement is found in all cases. Typical measurement uncertainties are at the level of $0.1\%$, which corresponds to 3.3 cm$^{-1}$ wavenumbers and is well within the range of spectroscopic accuracy for such molecular vibration problems. 

\section{Extension to Multiple Nuclear Degrees of Freedom}
\label{tensor}
\input{over-arching-theory}

\section{Conclusions}
\label{conclusions}
This paper has presented a rigorous theoretical framework and experimental demonstration for treating general chemical dynamics problems using quantum hardware. We have shown how molecular systems with anharmonic potential energy landscapes -- such as the ubiquitous hydrogen bond -- may be mapped to a system of controllable interacting quantum bits. We have performed a proof-of-principle demonstration of our mapping, finding excellent agreement between the observed hydrogen-bond dynamics and extracted spectra compared with those calculated on classical hardware. Finally, we discussed an outlook towards extending our approach to more complex chemical systems with multiple correlated nuclear degrees of freedom.

We remark that our experimental technique for extracting all energy eigenvalues of a many-body Hamiltonian has yielded higher accuracies (by nearly two orders-of-magnitude) than previous efforts using trapped-ion quantum systems \cite{senko2014coherent,jurcevic2015spectroscopy}. We note that this high accuracy is driven by the resilience of frequency information in our time-series data, despite accumulated quantum gate errors at the $\sim 10\%$ level. Our observations establish that current-generation, noisy quantum hardware can already serve as a precise computational resource for studying the spectral features of many-body Hamiltonians.

We have, for the first time, demonstrated that generalized nuclear dynamics can be modeled exactly using a qubit-based quantum processor. This is a marked departure from existing literature, where vibrations within the harmonic approximation are mapped to Bosonic systems. Building on the high-accuracy quantum simulations presented here, we will broaden our implementation to include multiple correlated nuclear degrees of freedom within the nuclear Hamiltonian to capture more realistic wavepacket trajectories across the potential energy surface and also allow for Boltzmann averaging over thermally-fluctuating donor-acceptor distances. Ultimately, we expect to compare and validate our quantum-computed results with experiments performing gas-phase spectroscopy, for which the Hamiltonian description of an isolated nuclear wavepacket is most appropriate.

\section{Acknowledgements}
The work of P.R., D.S., M.A.L-R., A.D., J.M.S., A.S., and S.S.I are supported by the U.S. National Science Foundation under award OMA-1936353. The QSCOUT open-access testbed is funded by the U.S. Department of Energy, Office of Science, Office of Advanced Scientific Computing Research Quantum Testbed Program. Sandia National Laboratories is a multimission laboratory managed and operated by National Technology \& Engineering Solutions of Sandia, LLC, a wholly owned subsidiary of Honeywell International Inc., for the U.S. Department of Energy’s National Nuclear Security Administration under contract DE-NA0003525. This paper describes objective technical results and analysis. Any subjective views or opinions that might be expressed in the paper do not necessarily represent the views of the U.S. Department of Energy or the United States Government.

\bibliography{main,dafsmooth,Quantum-Computing,PRX-refs,qwaimdrefs,SLO1,On-water-catalysis,water-clusters,math,bio,water-ox-Hbonds,SSI-molecular-fragmentation,admprefs,srini-water,carbocations,electronic-structure,materials,wavepacket,TN-paper,PES,htransfer}

\newpage

\onecolumngrid
\newpage
\begin{center}
\textbf{\large Supplementary Material for \\
``Quantum Computation of Hydrogen Bond Dynamics and Vibrational Spectra" \\~}
\end{center}
\twocolumngrid

\include{si}
\widetext

\end{document}

%% file: over-arching-theory.tex
The implementation of multi-dimensional molecular dynamics studies on quantum hardware is complicated by multiple factors. Molecules contain many correlated nuclear degrees of freedom, and each of these correlated nuclear degrees of freedom needs to be represented in some discretized basis. For ${\cal N}$ discretizations (or basis vectors) per nuclear degree of freedom and $N$ nuclear degrees of freedom, the complexity of information grows approximately as ${\cal N}^N$. As a result, quantum nuclear dynamics is thought to be an exponentially hard problem. In general, the Hamiltonian for such a system may be written as
\begin{align}
	{\cal{H}}(\vb{\bar{x}},\vb{\bar{x}}^\prime) &= {\cal K}(x_1,x_1^\prime) \otimes {\cal I}(x_2,x_2^\prime) \otimes \cdots \otimes {\cal I}(x_N,x_N^\prime) \nonumber \\ 
	&+ \sum_{i=2}^{N-1} {\cal I}(x_1,x_1^\prime) \otimes \cdots \otimes {\cal I}(x_{i-1},x_{i-1}^\prime) \otimes \nonumber \\ 
	&\phantom{+ } {\cal K}(x_i,x_i^\prime)
	\otimes {\cal I}(x_{i+1},x_{i+1}^\prime) \otimes \cdots \otimes {\cal I}(x_N,x_N^\prime) \nonumber \\ 
	&+ {\cal I}(x_1,x_1^\prime) \otimes \cdots \otimes {\cal I}(x_{N-1},x_{N-1}^\prime) \otimes {\cal K}(x_N,x_N^\prime) \nonumber \\ &+ {\cal V}(\vb{\bar{x}},\vb{\bar{x}}^\prime)
	\label{Eq:Hamil} 
\end{align}
where $\vb{\bar{x}}\equiv\qty{x_1,x_2,\ldots,x_N}$. While the kinetic energy part of the Hamiltonian, $\left\{ {\cal K}(x_i,x_i^\prime)\right\}$, is decoupled across all degrees of freedom and for each dimension is represented using Eq. (\ref{DAFfreeprop+derivative}), the potential energy surface correlates these dimensions and drives the computational complexity of the problem. Additionally, for quantum nuclear dynamics the term ${\cal V}(\vb{\bar{x}},\vb{\bar{x}}^\prime)$ is determined by the properties of electrons; when accurate electron correlation methods are employed, the computational cost for each electronic structure calculation may grow in a steeply algebraic manner. 
Furthermore, the number of terms needed to determine ${\cal V}(\vb{\bar{x}},\vb{\bar{x}}^\prime)$ may scale in the worst case as ${\cal N}^N$ thus rendering the quantum molecular dynamics problem, in conjunction with determination of the electronic potential energy surface, to be computationally intractable for large systems. This severely restricts the study of many problems at the forefront of biological, materials and atmospheric systems. 

For the process of quantum propagation, where either a quantum circuit is to be generated based on the time-evolution operator constructed from Eq. (\ref{Eq:Hamil}), or a controllable Ising-type quantum system is created that may be used to simulate the system depicted by Eq. (\ref{Eq:Hamil}), the very act of processing Eq. (\ref{Eq:Hamil}) requires, at the worst case, exponential resources. This can already be seen for the one-dimensional dynamics in DMANH$^+$: any oscillations in the N--N length will affect the nature of the double well potential that confines the shared hydrogen nucleus, requiring multiple iterative calculations to predict the spectroscopic behavior of the system. 

\input{Fig_MPS.tex}

Here we present a theoretical formalism for multi-dimensional quantum nuclear dynamics, the critical role played by the current experimental results, and the potential quantum advantage enabled by this framework. We begin by writing a general multi-dimensional quantum nuclear wavefunction as a tensor network. Specifically, in our case, we illustrate this by writing the wavefunction as a matrix product state (MPS) \cite{Ostlund1995-gi,Fannes1992-dd,Fannes1994-vz,Affleck1987-tx,Affleck1988-cj,Vidal2003-od,Schollwock2011-bl}:
\begin{align}
	\chi(\vb{\bar{x}};t) = \sum_{\bar{\boldsymbol\alpha}}^{\bar{\boldsymbol\eta}} \tensor*{\phi}{^{[1]}_{}^{x_1}_{\alpha_1}} \qty[\prod_{j=2}^{N\text{-} 1} \tensor*{\phi}{^{[j]}_{}^{x_j}_{\alpha_{j\text{-} 1}}_{\alpha_j}}] \tensor*{\phi}{^{[N]}_{}^{x_N}_{\alpha_{N\text{-} 1}}}.
	\label{Eq:MPS_WF} 
\end{align}
In other words, on the l.h.s. of Eq. \ref{Eq:MPS_WF}, the $N$-dimensional wavefunction $\chi(\vb{\bar{x}};t)$ is thought of as an order-$N$ tensor which is factorized into a product of order-2 and order-3 tensors $\phi^{[j]}$, each corresponding to one-dimensional functions of the coordinate $x_j$. The sum runs through the so-called entanglement variables $\bar{\boldsymbol\alpha} \equiv \qty{ \alpha_1, \alpha_2, \ldots, \alpha_{N-1} }$ with respective limits $\bar{\boldsymbol\eta} \equiv \qty{\eta_1,\eta_2,\ldots,\eta_{N-1}}$, commonly named entanglement dimension, bond dimension, or Schmidt rank. 
Equation (\ref{Eq:MPS_WF}) is graphically illustrated as a tensor network diagram in  Figure \ref{Fig:MPS_Diagram}. 
Using tensor networks, the time evolution operator for the Hamiltonian in Eq. (\ref{Eq:Hamil}) 
may also be written \cite{Schollwock2011-bl,Verstraete2004-qu,Pirvu2010-ll} as a tensor network operator:
\begin{align}
	{\cal{U}}(\vb{\bar{x}},\vb{\bar{x}}^\prime;t) = \sum_{\bar{\boldsymbol\beta}} \tensor*{{\cal{U}}}{^{[1]}_{}^{t; x_1,x_1^\prime}_{\beta_1}} \qty[\prod_{j=2}^{N\text{-} 1} \tensor*{{\cal{U}}}{^{[j]}_{}^{t; x_j,x_j^\prime}_{\beta_{j\text{-} 1}}_{\beta_j}}] \tensor*{{\cal{U}}}{^{[N]}_{}^{t; x_N,x_N^\prime}_{\beta_{N\text{-} 1}}},
	\label{Eq:MPS_WF-PropU} 
\end{align}
where the degree of entanglement captured within the time-evolution, designated by the parameter $\bar{\boldsymbol\beta}$ is evidently different from that in Eq. (\ref{Eq:MPS_WF}) and pertains to the physics captured within the Hamiltonian (and in fact ${\cal V}$ in Eq. (\ref{Eq:Hamil})) used to create such a time-evolution operator. Thus, we may equivalently write
\begin{align}
	{\cal{V}}(\vb{\bar{x}},\vb{\bar{x}}^\prime) = \delta_{\vb{\bar{x}},\vb{\bar{x}}^\prime} \sum_{\bar{\boldsymbol\beta}} \tensor*{{\cal{V}}}{^{[1]}_{}^{x_1,x_1^\prime}_{\beta_1}} \qty[\prod_{j=2}^{N\text{-} 1} \tensor*{{\cal{V}}}{^{[j]}_{}^{x_j,x_j^\prime}_{\beta_{j\text{-} 1}}_{\beta_j}}] \tensor*{{\cal{V}}}{^{[N]}_{}^{x_N,x_N^\prime}_{\beta_{N\text{-} 1}}},
	\label{Eq:MPS_WF-Prop} 
\end{align}
where it is assumed that the potential is diagonal in the coordinate representation $\ket{\bar{x}}$ used here.
Equation (\ref{Eq:MPS_WF-PropU}) yields the critical feature that each dimension can be independently propagated, on different quantum systems, and this aspect is pictorially represented using the purple and green edges in Figure \ref{Fig:MPS_tEvo}. In other words, Figure \ref{Fig:MPS_tEvo} depicts the full multi-dimensional time-evolution process as, 
\begin{align}
	\chi(\vb{\bar{x}};t) =& \int d\vb{\bar{x}}^\prime \; {\cal{U}}(\vb{\bar{x}},\vb{\bar{x}}^\prime;t) \; \chi(\vb{\bar{x}}^\prime;t=0) \nonumber \\ =& \sum_{\bar{\boldsymbol\alpha}}\sum_{\bar{\boldsymbol\beta}} \left[ \int dx_1^\prime \tensor*{{\cal{U}}}{^{[1]}_{}^{t; x_1,x_1^\prime}_{\beta_1}}\tensor*{\phi}{^{[1]}_{}^{x_1^\prime}_{\alpha_1}} \right] \nonumber \\ & \phantom{\sum_{\bar{\boldsymbol\alpha}}\sum_{\bar{\boldsymbol\beta}}} \qty[\prod_{j=2}^{N\text{-} 1} \int dx_j^\prime \tensor*{{\cal{U}}}{^{[j]}_{}^{t; x_j,x_j^\prime}_{\beta_{j\text{-} 1}}_{\beta_j}} \tensor*{\phi}{^{[j]}_{}^{x_j^\prime}_{\alpha_{j\text{-} 1}}_{\alpha_j}}] \nonumber \\ & \phantom{\sum_{\bar{\boldsymbol\alpha}}\sum_{\bar{\boldsymbol\beta}}} \left[ \int dx_N^\prime \tensor*{{\cal{U}}}{^{[N]}_{}^{t; x_N,x_N^\prime}_{\beta_{N\text{-} 1}}} \tensor*{\phi}{^{[N]}_{}^{x_N^\prime}_{\alpha_{N\text{-} 1}}}  \right] \nonumber \\ =& \sum_{\bar{\boldsymbol\alpha},\bar{\boldsymbol\beta}} \tensor*{\phi}{^{[1]}_{}^{t; x_1}_{\beta_1\alpha_1}} \qty[\prod_{j=2}^{N\text{-} 1} \tensor*{\phi}{^{[j]}_{}^{t; x_j}_{\beta_{j\text{-} 1}}_{\beta_j}_{\alpha_{j\text{-} 1}}_{\alpha_j}}] \tensor*{\phi}{^{[N]}_{}^{t; x_N}_{\beta_{N\text{-} 1}}_{\alpha_{N\text{-} 1}}}.
	\label{Eq:MPS_Propagated-state} 
\end{align}
\input{Fig_MPS_tEvo.tex}
\noindent where the time evolution for each component, for example, depicted by the purple and green edges in Figure \ref{Fig:MPS_tEvo}:
\begin{align}
\tensor*{\phi}{^{[j]}_{}^{t; x_j}_{\beta_{j\text{-} 1}}_{\beta_j}_{\alpha_{j\text{-} 1}}_{\alpha_j}} \equiv \int dx_j'\,\tensor*{{\cal{U}}}{^{[j]}_{}^{t; x_j,x_j^\prime}_{\beta_{j\text{-} 1}}_{\beta_j}} \tensor*{\phi}{^{[j]}_{}^{x_j^\prime}_{\alpha_{j\text{-} 1}}_{\alpha_j}}
\label{Eq:MPS_Propagated-state-1D}
\end{align}
may be carried out on separate quantum simulators. Equation (\ref{Eq:MPS_Propagated-state}) is completely represented in Figure \ref{Fig:MPS_tEvo}. The top row here is the MPS state in Eq. (\ref{Eq:MPS_WF}), as evidenced through comparison with Figure \ref{Fig:MPS_Diagram}. The second row (green) represents the propagator in Eq. (\ref{Eq:MPS_WF-Prop}) with entanglement dimensions $\bar{\boldsymbol\beta}$; the connected green and purple lines indicate the integration over the primed variables in Eqs. (\ref{Eq:MPS_Propagated-state}) and (\ref{Eq:MPS_Propagated-state-1D}). As noted, each such integration (or time-evolution) may be mapped on to a separate quantum system or quantum processor. Finally the bottom (blue) row of Figure \ref{Fig:MPS_Diagram} shows the right side of Eq. (\ref{Eq:MPS_Propagated-state}) where the entanglement variables are now double indexed as $\bar{\boldsymbol\alpha},\bar{\boldsymbol\beta}$. 

Given the structure and sparsity of the system Hamiltonian, it is critical to assess the number of parameters in the unitary evolution depicted by Eqs. (\ref{Eq:MPS_WF-PropU}) and (\ref{Eq:MPS_WF-Prop}) since these will eventually determine the nature of the quantum circuit, the associated depth, and complexity. At the outset, let us emphasize that the number of parameters within the direct unitary evolution of Eq. (\ref{Eq:Hamil}) could be as many as ${\cal O} \left( {\cal N}^N\right)$, as already noted above. This implies that, when using $n$ qubits per nuclear dimension, the number of parameters in Eq. (\ref{Eq:Hamil}) is $2^{nN}$ in the worst case. 

However, when Eqs. (\ref{Eq:MPS_WF-PropU}) and (\ref{Eq:MPS_WF-Prop}) are used for quantum propagation, one only needs of the order of $\left( {\cal N}*N\right)$, or $2^{n}N$ parameters inside the same set of $n*N$ qubit quantum system. This is because, for most chemical systems the number of product vectors in Eq. (\ref{Eq:MPS_WF-Prop}) is expected to be  small number as compared to the total number of nuclear basis functions, ${\cal N}^N$. This result appears from  the ``area law of entanglement entropy'' \cite{tensor_network}, where it is thought to be the case that the extent of entanglement in most physical systems, as depicted by the number of product functions in Eq. (\ref{Eq:MPS_WF-Prop}), scales as the area, rather than the volume of the Hilbert space. In other words, the number of non-zero terms in Eq. (\ref{Eq:MPS_WF-Prop}) is of order $\left( {\cal N}*N\right)$ times the number of terms in the summation in Eq. (\ref{Eq:MPS_WF-Prop}). This last part is expected to be small compared to both ${\cal N}$ and $N$, for large enough values of both parameters, based on the area law of entanglement entropy. This is a major advantage for the current formalism, and future experimental implementations will actively probe the limits of this apparent theoretical quantum advantage that promises an exponential circuit depth reduction of the order of $2^{nN}/(2^{n}N)$.

Thus, the tensor network formalism allows the overall unitary propagator to be factorized into separate pieces shown as the green blocks in Figure \ref{Fig:MPS_tEvo}; the action of each of which may be constructed in parallel as is seen from the above equations. 
This critical feature when combined with the one-dimensional propagation algorithm presented here and in Ref. \onlinecite{saha2021mapping}, yields a general approach to treat the classically intractable multi-dimensional quantum nuclear dynamics problem on a family of quantum systems. 

%% file: Fig_MPS.tex
\definecolor{bblue}{rgb}{0.19, 0.55, 0.91}
\definecolor{green}{rgb}{0.0, 0.65, 0.58}
\definecolor{purple}{rgb}{0.6, 0.4, 0.8}
\definecolor{orange}{rgb}{0.83, 0.4, 0.32}

\begin{figure}[t!]
    \centering
    \tikzstyle{ellps} = [ellipse, very thick, inner sep=0mm, draw=purple, fill=purple!20,minimum width=3.5cm,minimum height=0.7cm,align=center]

    \tikzstyle{phi} = [circle, very thick, minimum size=0.8cm, inner sep=0.2mm, draw=purple, fill=purple!20]

    \tikzset{arw/.style={-Stealth, thick}}

    \begin{tikzpicture}[>=latex,node distance=3cm,
        every edge quote/.style={font=\fontsize{7}{1}},
        every node/.style={font=\fontsize{8}{1}}
        ]
        \node (psi) [ellps] {$\chi$};
        \node (xindex) [below=3mm of psi, inner sep=0mm]{};
        \node (x1) [left=10.5mm of xindex] {};
        \node (x2) [right=3mm of x1] {};
        \node (cdots) [right=3mm of x2] {$\cdots$};
        \node (xN_1) [right=3mm of cdots] {};
        \node (xN) [right=3mm of xN_1] {};
        
        \node (mps) [below=10mm of psi,text=purple,inner sep=0.2cm]{MPS};
        \node (mps0) [above=4mm of mps] {};
        \node (phidots) [below=3mm of mps, minimum size=8mm]{$\cdots$};
        \node (phi2) [phi, left=6mm of phidots,font=\fontsize{7}{1}]{$\phi^{[2]}$};
        \node (phi1) [phi, left=9mm of phi2,font=\fontsize{7}{1}]{$\phi^{[1]}$};
        \node (phiN_1) [phi, right=6mm of phidots,font=\fontsize{7}{1}]{$\phi^{[N\text{-}1]}$};
        \node (phiN) [phi, right=9mm of phiN_1,font=\fontsize{7}{1}]{$\phi^{[N]}$};
        
        \node (px1) [below=2mm of phi1]{};
        \node (px2) [below=2mm of phi2]{};
        \node (pxN_1) [below=2mm of phiN_1]{};
        \node (pxN) [below=2mm of phiN]{};

        \begin{pgfonlayer}{background}
            \node (ix1) [above=4mm of x1]{};
            \node (ix2) [above=4mm of x2]{};
            \node (ixN_1) [above=4mm of xN_1]{};
            \node (ixN) [above=4mm of xN]{};
            
            \path[-]
        		(ix1) edge[very thick, purple] node[below,black,yshift=-1.5mm] {$_{x_1}$} (x1)
        		(ix2) edge[very thick, purple] node[below,black,yshift=-1.5mm] {$_{x_2}$} (x2)
        		(ixN_1) edge[very thick, purple] node[below,black,yshift=-1.5mm] {$_{x_{N\text{-}1}}$} (xN_1)
        		(ixN) edge[very thick, purple] node[below,black,yshift=-1.5mm] {$_{x_N}$} (xN)
        		;
        \end{pgfonlayer}

        \path[-]
            (phi1) edge[very thick, purple] node[above,black] {$_{\alpha_1}$} (phi2)
            (phi2) edge[very thick, purple] node[above,black] {$_{\alpha_2}$} (phidots)
            (phidots) edge[very thick, purple] node[above,black,xshift=-1mm] {$_{\alpha_{N\text{-}2}}$}  (phiN_1)
            (phiN_1) edge[very thick, purple] node[above,black] {$_{\alpha_{N\text{-}1}}$}  (phiN)
            (phi1) edge[very thick, purple] node[left,black] {$_{x_1}$} (px1)
            (phi2) edge[very thick, purple] node[left,black] {$_{x_2}$} (px2)
            (phiN_1) edge[very thick, purple] node[left,black] {$_{x_{N\text{-}1}}$} (pxN_1)
            (phiN) edge[very thick, purple] node[left,black] {$_{x_N}$} (pxN)           
            ;
           
           \draw[arw,thick,draw=purple] (mps0.south) -| (mps.north);
           \draw[arw,thick,draw=purple,rounded corners=1mm,dashed] (mps.west) -| (phi1.north);
           \draw[arw,thick,draw=purple,rounded corners=1mm,dashed] (mps.west) -| (phi2.north);
           \draw[arw,thick,draw=purple,rounded corners=1mm,dashed] (mps.east) -| (phiN_1.north);
           \draw[arw,thick,draw=purple,rounded corners=1mm,dashed] (mps.east) -| (phiN.north);
           
    \end{tikzpicture}

    \caption{Tensor network diagram of the MPS representation of the wave function $\chi(\vb{\bar{x}};t)$. (See Eq. (\ref{Eq:MPS_WF}).) In the discrete position basis, the $N$-dimensional wave function is pictured as an $N$-order tensor, which is factorized as a product of order-2 and order-3 tensors $\phi^{[k]}$.}
    \label{Fig:MPS_Diagram}
\end{figure}
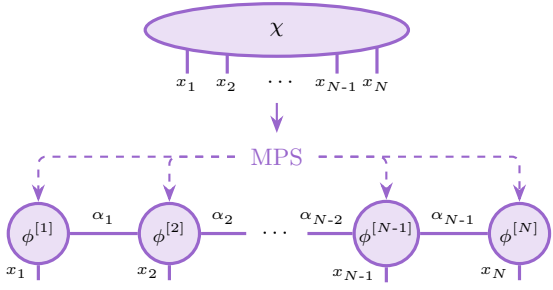

%% file: Fig_MPS_tEvo.tex
\definecolor{bblue}{rgb}{0.19, 0.55, 0.91}
\definecolor{green}{rgb}{0.0, 0.65, 0.58}
\definecolor{purple}{rgb}{0.6, 0.4, 0.8}
\definecolor{orange}{rgb}{0.83, 0.4, 0.32}

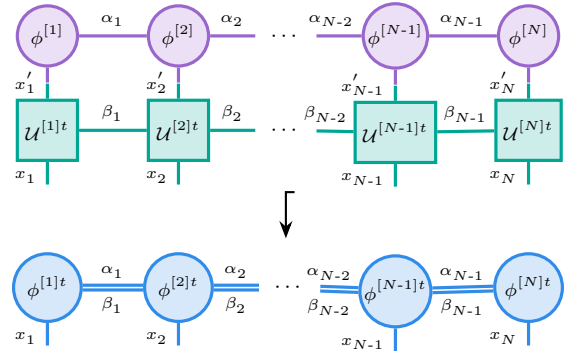
\begin{figure}[ht!]
    \centering
    \tikzstyle{ellps} = [ellipse, very thick, inner sep=0mm, draw=purple, fill=purple!20,minimum width=3.5cm,minimum height=0.7cm,align=center]

    \tikzstyle{phi} = [circle, very thick, minimum size=0.8cm, inner sep=0.2mm, draw=purple, fill=purple!20]
    
    \tikzstyle{phinew} = [circle, very thick, minimum size=0.9cm, inner sep=0.2mm, draw=bblue, fill=bblue!20]
    
    \tikzstyle{U} = [rectangle, very thick, minimum size=0.8cm, inner sep=0.1cm,draw=green, fill=green!20]
    
    \tikzset{arw/.style={-Stealth, thick}}

    \begin{tikzpicture}[>=latex,node distance=3cm,
        every edge quote/.style={font=\fontsize{7}{1}},
        every node/.style={font=\fontsize{8}{1}}
        ]
		\node (pphi1) [phi,font=\fontsize{7}{1}]{$\phi^{[1]}$};
        \node (pphi2) [phi, right=9mm of pphi1,font=\fontsize{7}{1}]{$\phi^{[2]}$};
        \node (pphidots) [right=6mm of pphi2, minimum size=8mm]{$\cdots$};
        \node (pphiN_1) [phi, right=6mm of pphidots,font=\fontsize{7}{1}]{$\phi^{[N\text{-}1]}$};
        \node (pphiN) [phi, right=9mm of pphiN_1,font=\fontsize{7}{1}]{$\phi^{[N]}$};        
        \node (ppx1) [below=2mm of pphi1, minimum size=0mm, inner sep=0mm]{};
        \node (ppx2) [below=2mm of pphi2, minimum size=0mm, inner sep=0mm]{};
        \node (ppxN_1) [below=2mm of pphiN_1, minimum size=0mm, inner sep=0mm]{};
        \node (ppxN) [below=2mm of pphiN, minimum size=0mm, inner sep=0mm]{};
        
        \node (u1) [U, below=2mm of ppx1, font=\fontsize{7}{1}]{$\mathcal{U}^{[1]t}$};
        \node (u2) [U, below=2mm of ppx2, font=\fontsize{7}{1}]{$\mathcal{U}^{[2]t}$};
        \node (udots) [right=6mm of u2, minimum size=8mm]{$\cdots$};
        \node (uN_1) [U, below=2mm of ppxN_1, font=\fontsize{7}{1}]{$\mathcal{U}^{[N\text{-}1]t}$};
        \node (uN) [U, below=2mm of ppxN, font=\fontsize{7}{1}]{$\mathcal{U}^{[N]t}$};        
        \node (ux1) [below=3mm of u1, minimum size=0mm, inner sep=0mm]{};
        \node (ux2) [below=3mm of u2, minimum size=0mm, inner sep=0mm]{};
        \node (uxN_1) [below=3mm of uN_1, minimum size=0mm, inner sep=0mm]{};
        \node (uxN) [below=3mm of uN, minimum size=0mm, inner sep=0mm]{};
        
        \node (eql0sgn) [below=3mm of udots]{};
        \node (eqlsgn) [below=5mm of eql0sgn]{};  
        
        \node (phit1) [phinew, below=12mm of u1, font=\fontsize{7}{1}]{$\phi^{[1]t}$};
        \node (phit2) [phinew, below=12mm of u2, font=\fontsize{7}{1}]{$\phi^{[2]t}$};
        \node (phitdots) [right=6mm of phit2, minimum size=8mm]{$\cdots$};
        \node (phitN_1) [phinew, below=12mm of uN_1, font=\fontsize{7}{1}]{$\phi^{[N\text{-}1]t}$};
        \node (phitN) [phinew, below=12mm of uN, font=\fontsize{7}{1}]{$\phi^{[N]t}$};        
        \node (phitx1) [below=3mm of phit1, minimum size=0mm, inner sep=0mm]{};
        \node (phitx2) [below=3mm of phit2, minimum size=0mm, inner sep=0mm]{};
        \node (phitxN_1) [below=3mm of phitN_1, minimum size=0mm, inner sep=0mm]{};
        \node (phitxN) [below=3mm of phitN, minimum size=0mm, inner sep=0mm]{};

        \draw[arw, to path={-| (\tikztotarget)}] (eql0sgn) edge (eqlsgn);
        
        \path[-]
            
           
            
            (pphi1) edge[very thick, purple] node[above,black] {$_{\alpha_1}$} (pphi2)
            (pphi2) edge[very thick, purple] node[above,black] {$_{\alpha_2}$} (pphidots)
            (pphidots) edge[very thick, purple] node[above,black,xshift=-1mm] {$_{\alpha_{N\text{-}2}}$}  (pphiN_1)
            (pphiN_1) edge[very thick, purple] node[above,black] {$_{\alpha_{N\text{-}1}}$}  (pphiN)
            (pphi1) edge[very thick, purple] node[left,black,yshift=-1mm] {$_{x'_1}$} (ppx1)
            (pphi2) edge[very thick, purple] node[left,black,yshift=-1mm] {$_{x'_2}$} (ppx2)
            (pphiN_1) edge[very thick, purple] node[left,black,yshift=-1mm] {$_{x'_{N\text{-}1}}$} (ppxN_1)
            (pphiN) edge[very thick, purple] node[left,black,yshift=-1mm] {$_{x'_N}$} (ppxN)
            
            (u1) edge[very thick, green] node[above,black] {$_{\beta_1}$} (u2)
            (u2) edge[very thick, green] node[above,black] {$_{\beta_2}$} (udots)
            (udots) edge[very thick, green] node[above,black,xshift=-1.2mm] {$_{\beta_{N\text{-}2}}$} (uN_1)
            (uN_1) edge[very thick, green] node[above,black] {$_{\beta_{N\text{-}1}}$} (uN)
            (u1) edge[very thick, green] (ppx1)
            (u1) edge[very thick, green] node[left,black,yshift=-0.5mm] {$_{x_1}$}(ux1)
            (u2) edge[very thick, green] (ppx2)
            (u2) edge[very thick, green] node[left,black,yshift=-0.5mm] {$_{x_2}$}(ux2)
            (uN_1) edge[very thick, green] (ppxN_1)
            (uN_1) edge[very thick, green] node[left,black,yshift=-0.5mm] {$_{x_{N\text{-}1}}$}(uxN_1)
            (uN) edge[very thick, green] (ppxN)
            (uN) edge[very thick, green] node[left,black,yshift=-0.5mm] {$_{x_N}$}(uxN)
            
            (phit1) edge[double,very thick, bblue] node[above,black] {$_{\alpha_1}$} node[below,black] {$_{\beta_1}$} (phit2)
            (phit2) edge[double,very thick, bblue] node[above,black] {$_{\alpha_2}$} node[below,black] {$_{\beta_2}$} (phitdots)
            (phitdots) edge[double,very thick, bblue] node[above,black,xshift=-1.3mm] {$_{\alpha_{N\text{-}2}}$} node[below,black,xshift=-1.3mm] {$_{\beta_{N\text{-}2}}$} (phitN_1)
            (phitN_1) edge[double,very thick, bblue] node[above,black] {$_{\alpha_{N\text{-}1}}$} node[below,black] {$_{\beta_{N\text{-}1}}$} (phitN)
            (phit1) edge[very thick, bblue] node[left,black,yshift=-0.5mm] {$_{x_1}$}(phitx1)
            (phit2) edge[very thick, bblue] node[left,black,yshift=-0.5mm] {$_{x_2}$}(phitx2)
            (phitN_1) edge[very thick, bblue] node[left,black,yshift=-0.5mm] {$_{x_{N\text{-}1}}$}(phitxN_1)
            (phitN) edge[very thick, bblue] node[left,black,yshift=-0.5mm] {$_{x_N}$}(phitxN)
            ;

    \end{tikzpicture}

    \caption{Action of the time evolution operator $\mathcal{U}(\vb{\bar{x}},\vb{\bar{x}'};t)$ onto the MPS wave function $\chi(\vb{\bar{x}};t=0)$ shown in Eq. (\ref{Eq:MPS_Propagated-state}). }
    \label{Fig:MPS_tEvo}
\end{figure}

%% file: si.tex



\title{Supplementary Material for \\
``Quantum Computation of Hydrogen Bond Dynamics and Vibrational Spectra"}

\maketitle

\noindent\textbf{DMANH$^+$ Molecule} -- 
The specific intra-molecular proton transfer problem 
considered here is that in the protonated 1,8-bis(dimethylamino)naphthalene (DMANH$^+$) system. The DMAN molecule has an extremely large proton affinity of $242$ kcal/mol \cite{lias1984j}, with DMANH$^+$  $pK_{a}$ value in the range $12.1-12.3$ \cite{perrin1972dissociation}. The NHN$^+$ hydrogen bond in DMANH$^+$ has been frequently studied as a model for short, low-barrier hydrogen bonds that have a role in certain enzyme-catalyzed reactions \cite{SSHB-NielsonReview,Sshb-Enzymes}. In solution, the shared proton delocalization in DMANH$^+$ is controlled by a  
low-barrier symmetric double-well potential, with barrier height being influenced by solvent and temperature \cite{Jeremy-ref1,Jeremy-ref2}. 
In this publication, the shared proton reduced dimensional potential for the most significant donor-acceptor distance, that is N-N distance equal to 2.53 \AA, is used to perform quantum dynamics studies on an ion-lattice quantum computer.
We treat the shared proton stretch dimension within the 
Born-Oppenheimer limit. The nuclear Hamiltonian is determined by the ground electronic state 
potential energy surface. 

\bigskip

\noindent
\textbf{Quantum Computation using the QSCOUT ion-trap device} -- 
The experiment described here was performed on the room temperature Quantum Scientific Computing Open User Testbed (QSCOUT) system~\cite{clark2021engineering}. In this work, we used two $^{171}$Yb$^+$ ions trapped $70~\mu$m off the surface of an HOA-2.1 trap~\cite{maunz2016high} with radial secular frequencies , $\nicefrac{\omega}{2 \pi}$, of 2.0~MHz and 2.4~MHz. The axial frequency is set to match our individual addressing and imaging fiber spacing of about $4.5~\mu$m. As previously mentioned, the ions are manipulated between the $^2$S$_{1/2}\ket{F=0,m_F=0}$ and $\ket{F=1,m_F=0}$ hyperfine states, denoted $\ket{0}$ and $\ket{1}$, respectively \cite{olmschenk2007manipulation}. We use Doppler cooling and resolved sideband cooling to cool the ions to near the ground state then prepare them in the $\ket{0}$ state using optical pumping. 

The coherent operations are performed using two counterpropagating 355~nm Raman beams from a single laser. These drive a two-photon transition~\cite{wineland1998experimental, leibfried2003raman} via a virtual state, 33~THz detuned from the $^2$P$_{1/2}$ level. The global beam has an elliptical $1/e^2$ beam waist of $10.9~\mu$m vertically and $182.8~\mu$m horizontally, such that scattering off the surface is minimized but the ions are illuminated with near uniform intensity. The individual addressing beams are created using a 32-channel acousto-optic modulator (AOM) with integrated optics~\cite{Harris2020aom} for creating 32 parallel and individually controllable beams. We use custom optics to create an elliptical profile at the ion from each of these beams with a vertical waist of $8~\mu$m, also to minimize scatter off the surface, and a horizontal beam waist of $0.8~\mu$m to achieve a $4.5~\mu$m ion spacing. 

The Jaqal code that defines the gate sequences and rotation angles is fed directly into the experimental machine. 
This work used a combination of single qubit gates and two qubit CNOT gates. The single qubit rotations, $R(\theta,\phi)$, where $\theta$ is the rotation angle and $\phi$ is the phase defined in the Jaqal code, are generated using only the individual addressing beam in co-propagating configuration by applying both Raman tones. The $R_z$ gate is performed in software using a phase advance. 
The single qubit gate times are approximately $15~\mu$s and single qubit gate fidelities are about 99.5(3)\%.

The CNOT gate is a M\o lmer-S\o rensen (MS) interaction between a series of single qubit rotations. We perform the MS gate by applying two frequency tones on the individual addressing beam corresponding to each ion and a single frequency tone on the global beam. The combination of the frequencies results in a pulse capable of driving the MS interaction. 
MS gate times are approximately $200~\mu$s and fidelities are 97(1)\%. 
We create a CNOT gate using an MS gate with a series of single qubit wrapper gates which give a combined CNOT gate fidelity of 96.5(1.5)\%. 

We use a high numerical aperture imaging system to reduce the detection time and increase readout fidelity~\cite{noek2013imaging}. The collected light is coupled into a multimode fiber array, such that each ion location is aligned to a different fiber~\cite{clark2021engineering}. Each fiber is connected to a dedicated photomultiplier tube for counting the photons, thus allowing for simultaneous and distinguishable detection between the two ions.  An elliptical $370~$nm laser beam that is resonant to the $^2$S$_{1/2} \ket{F=1,m_F=0}\rightarrow^2$P$_{1/2}\ket{F=0,m_F=0}$ transition illuminates both ions with comparable intensity. We apply the resonant light for $350~\mu$s, ions in state $\ket{1}$ are detected as bright with 99(0.3)\% fidelity. 

\bigskip

\noindent
\textbf{Mapping from Ion Trap to Molecule} --
After allowing for time evolution in each of the sub-blocks and detecting the ion qubit states, the measurement outcomes from the ion-trap system must be re-mapped back to the proton grid basis. For specificity in this section, we will focus on the mapping to the first proton lattice site $\ket{x_1}$; the re-mapping to all other lattice sites follows an identical procedure. 

From the map definition in Eq. 5 of the main text, the probability amplitude $c_1$ for the shared-proton wavepacket on lattice site $\ket{x_1}$ is given by
\begin{equation}
    c_1 = \frac{1}{\sqrt{2}}(c_{\tilde{1}} + c_{\tilde{8}})
\end{equation}
where the $c_{\tilde{n}}$ indicate probability amplitudes in the ion-trap basis (see Eqs. 5-6 in the main text). The overall lattice site probability $P(\ket{x_1})$ is then
\begin{align}
\label{remap-prob}
    P(\ket{x_1}) &= |c_1|^2 = \frac{1}{2}(|c_{\tilde{1}}|^2 + |c_{\tilde{8}}|^2 + 2 \text{Re}[c_{\tilde{1}}^* c_{\tilde{8}}]) \\
    &= \frac{1}{2}(P(\ket{\tilde{1}}) + P(\ket{\tilde{8}})+2 \text{Re}[c_{\tilde{1}}^* c_{\tilde{8}}])
\end{align}
where $P(\ket{\tilde{1}})$ and $P(\ket{\tilde{8}})$ are the probabilities of measuring the first state in the upper sub-block, and the last state in the lower sub-block, respectively.

To map the cross-term, we write the complex probability amplitude $c_{\tilde{n}}=A_{\tilde{n}} e^{i\phi_{\tilde{n}}} = \sqrt{P(\ket{\tilde{n}})} e^{i\phi_{\tilde{n}}}$, such that
\begin{equation}
    \text{Re}[c_{\tilde{1}}^* c_{\tilde{8}}] = \sqrt{P(\ket{\tilde{1}}) P(\ket{\tilde{8}})} \cos(\phi_{\tilde{8}}-\phi_{\tilde{1}})
\end{equation}
Finally, using that the phase accrued at time $t$ can be written to first order as $\phi_{\tilde{n}} = \tilde{\mathcal{H}}^{Mol}_{nn}t/\hbar$, the full shared-proton probability (Eq. \ref{remap-prob}) may be constructed using only ion-trap measurement probabilities and the original molecular Hamiltonian.

\bigskip

\noindent\textbf{Vibrational Spectra from time-correlation functions} --
We examine our quantum circuit based dynamics algorithm by simulating 
the reduced dimensional shared proton vibrational dynamics in the 1,8-bis(dimethylamino) naphthalene (DMANH$^+$)  molecular system on an ion-trap quantum computing system. We compare our results with results from classical simulation of the quantum dynamics. 
As a result, for the first time, we present, in this paper, vibrational properties determined from quantum computing experiments. The approach used to achieve this is as follows.

The time-dependent density matrix in an eigenstate representation is given by,
\begin{eqnarray}
    \rho(t) &=& \ket{\chi(t)}\bra{\chi(t)} = 
    e^{-\imath {\cal H}t /\hbar}
    \ket{\chi(0)}\bra{\chi(0)}
    e^{\imath {\cal H}t/{\hbar}}
    \nonumber \\
    &=& \sum_{i,j}c_{i}(0)c_{j}^{*}(0) 
    e^{-\imath{\cal H}t/{\hbar}}
    \ket{\phi_{i}}\bra{\phi_{j}}
    e^{\imath {\cal H}t/{\hbar}}
    \nonumber \\
    &=& \sum_{i,j}c_{i}(0)c_{j}^{*}(0) 
    e^{\imath (E_{i}-E_{j}) t/{\hbar}}
    \ket{\phi_{i}}
    \bra{\phi_{j}},
    \label{rhodefn}
\end{eqnarray}
and is used to construct the Fourier transform of the density-density time auto-correlation function, ${\text{Tr}}[\rho(0)\rho(t)]$ as
\begin{widetext}
\begin{eqnarray}
    \int_{-\infty}^{+\infty} dt 
    e^{{\imath \omega t}} \; {\text{Tr}}[\rho(0)\rho(t)] &=& \int_{-\infty}^{+\infty} dt e^{{\imath \omega t}}
    \; {\text{Tr}}\left[\ket{\chi(0)}\bra{\chi(0)}\sum_{i,j}c_{i}(0)c_{j}^{*}(0) 
    e^{\imath (E_{i}-E_{j}) t/{\hbar}}
    \ket{\phi_{i}}\bra{\phi_{j}}\right] \nonumber \\ 
    &=& \sum_{i,j}|c_{i}(0)|^{2}|c_{j}(0)|^{2} \; \delta \left(\omega-(E_{i}-E_{j})/\hbar \right)
    \label{Density-timecorrelation-FT}
\end{eqnarray}
\end{widetext}
which yields the difference in eigenenergies, weighted by contributions from the initial state, much like a standard experimental result. Equation 10 in the main text has a similar form as Eq. \ref{Density-timecorrelation-FT} and also depends on the difference in eigenenergies, but differs in intensities. To gauge the relation between the intensities in Eqs. (10) and (\ref{Density-timecorrelation-FT}), one may rewrite Eq. (\ref{Density-timecorrelation-FT}) using the convolution theorem \cite{Numerical-Recipes} as
\begin{align}
    & \int dx dx^\prime \int_{-\infty}^{+\infty} dt dt^\prime e^{{\imath \omega (t-t^\prime)}} \;\rho(x;x^\prime;t)\rho(x,x^\prime;t^\prime) = \nonumber \\
    & \phantom{\int dx dx^\prime} \int dx dx^\prime {\left\{ \left\vert \int_{-\infty}^{+\infty} dt  e^{{\imath \omega t}} \;\rho(x;x^\prime;t) \right\vert \right\}}^2
    \label{Density-timecorrelation-FT-1}
\end{align}
Equation 10 in the main text is simply the diagonal ($x=x^\prime$) portion of the term inside curly brackets $\left\{ \cdots \right\}$ on the rights side of  Eq. (\ref{Density-timecorrelation-FT-1}). For example, the right side of Eq. (\ref{Density-timecorrelation-FT-1}) may be written as 
\begin{align}
    & \int dx  {\left\vert \int_{-\infty}^{+\infty} dt  e^{{\imath \omega t}} \;\rho(x;x;t) \right\vert}^2
     + \nonumber \\ & \int_{x\neq x^\prime} dx dx^\prime {\left\vert \int_{-\infty}^{+\infty} dt  e^{{\imath \omega t}} \;\rho(x;x^\prime;t) \right\vert}^2
\end{align}
and using Eq. (\ref{rhodefn}), 
further reduced to, 
\begin{widetext}
\begin{align}
    & \int dx  {
    \left\{ 
    \left\vert 
    \int_{-\infty}^{+\infty} dt  e^{{\imath \omega t}} \;
    \sum_{i,j}c_{i}(0)c_{j}^{*}(0) 
    e^{\imath (E_{i}-E_{j}) t/{\hbar}}
    {\phi_{i}(x)}
    {\phi_{j}(x)} 
    \right\vert 
    \right\}
    }^2 +
    \nonumber \\ & 
    \int_{x\neq x^\prime} dx dx^\prime \left\vert \int_{-\infty}^{+\infty} dt  e^{{\imath \omega t}} 
    \sum_{i,j}c_{i}(0)c_{j}^{*}(0) 
    e^{\imath (E_{i}-E_{j}) t/{\hbar}}
    {\phi_{i}(x)}
    {\phi_{j}(x^\prime)} \right\vert^2
    \label{Density-timecorrelation-FT-2} 
\end{align}
\end{widetext}
Note that the term inside curly brackets $\left\{ \cdots \right\}$ on the first line of Eq. (\ref{Density-timecorrelation-FT-2}) is identical to the left side of Eq. 10 in the main text. Thus by measuring Eq. 10, we have a direct gauge on the standard time-correlation function described here, and thus we provide a new approach to determine spectroscopic features in complex systems using quantum computing platforms. 

\bigskip

\noindent\textbf{Determination of Ising Couplings and Local Fields for Molecular Hamiltonians} -- As is clear from the above discussion, Eq. 16 in the main text will need to be propagated on a ``tailored'' quantum system, and all such quantum systems are to be harnessed together as represented within Eq. 15 of the main text. In this paper we have experimentally verified that the computation in Eq. 16 can be performed on an ion-trap, albeit on a small number of qubits. Here we provide the critical steps needed to perform the one-dimensional operation in Eq. 16 on a large number of qubits. 

First the computational basis states represented within an ion trap are separated into odd, $\left\{ {\bf S^+}^{2n-1} \ket{\downarrow \downarrow \cdots}\right\}$, and even, $\left\{ {\bf S^+}^{2n} \ket{\downarrow \downarrow \cdots}\right\}$, spans of the total spin raising operators. Following this, the ion-lattice Hamiltonian, 
\begin{eqnarray}
{\cal H}_{IT} &=& \sum_{\gamma}
\sum_{i =1}^{N-1}
\sum_{j>i}^N
J_{ij}^{\gamma}\sigma_{i}^{\gamma}\sigma_{j}^{\gamma} +\sum_{\gamma}\sum_{i=1}^{N}B_{i}^{\gamma} \sigma_{i}^{\gamma}
\label{HIT-gen}  
\end{eqnarray}
where $\gamma \in {(x,y,z)}$, and $N$ is the number of qubits (or ion-sites). 
The quantities $\left\{ \sigma_i^\gamma \right\}$ are the Pauli spin operators acting on the $i^{th}$ lattice site along the $\gamma$-direction of the Bloch sphere. Equation (\ref{HIT-gen}) yields a block structure where the states within odd or even permuted block, $\bra{\downarrow \downarrow \cdots} {\bf S^{+^{2n}}} {\cal H}_{IT} {\bf S^{+^{2n}}} \ket{\downarrow \downarrow \cdots}$ and $\bra{\downarrow \downarrow \cdots} {\bf S^{+^{2n-1}}} {\cal H}_{IT} {\bf S^{+^{2n-1}}} \ket{\downarrow \downarrow \cdots}$, are coupled through the $J^{x}$ and $J^{y}$ parameters, whereas the states across these blocks, $\bra{\downarrow \downarrow \cdots} {\bf S^{+^{2n}}} {\cal H}_{IT} {\bf S^{+^{2n-1}}} \ket{\downarrow \downarrow \cdots}$ and complex conjugate, are coupled through $B^{x}$ and $B^{y}$. That is, the matrix form of Eq. (\ref{HIT-gen}) in the permuted basis stated above takes the block form for $N$-qubits as
\begin{align}
    \mathbf{{H}_{N}} 
    &= \begin{bmatrix}
    \mathbf{H_{N}^{D1}} & \mathbf{B_{N}} \\
    \mathbf{B_{N}^{\dagger}} & \mathbf{H_{N}^{D2}} \\
    \end{bmatrix}
\end{align}
or, recursively as, 
\begin{widetext}
\begin{align}
    \mathbf{{H}_{N}} 
    &= 
    \left[ \begin{array}{@{}c|c@{}} 
    \begin{matrix}
        \substack{\mathbf{{H}_{N-1}^{D1}} + B_{N}^{z}\mathbf{I_{2^{N-2}}} \\ +\mathbf{J_{z,N}^{1}}} & \mathbf{J_{xy,N}^{1}}\\
        \mathbf{J_{xy,N}^{1}}^{\top} & \substack{\mathbf{{H}_{N-1}^{D2}} - B_{N}^{z}\mathbf{I_{2^{N-2}}}  \\+ \mathbf{J_{z,N}^{1}}}
    \end{matrix}
    & \begin{matrix}
        \mathbf{B_{N-1}} & \left(B_{N}^{x} - \imath B_{N}^{y} \right)\mathbf{I_{2^{N-2}}}\\
        \left(B_{N}^{x} + \imath B_{N}^{y} \right)\mathbf{I_{2^{N-2}}} & \mathbf{B_{N-1}}
    \end{matrix}\\
    \hline
    \begin{matrix}
        \mathbf{B_{N-1}^{\dagger}} & \left(B_{N}^{x} - \imath B_{N}^{y} \right)\mathbf{I_{2^{N-2}}}\\
        \left(B_{N}^{x} + \imath B_{N}^{y} \right)\mathbf{I_{2^{N-2}}} & \mathbf{B_{N-1}^{\dagger}}
    \end{matrix}
    & \begin{matrix}
        \substack{\mathbf{{H}_{N-1}^{D2}} + B_{N}^{z}\mathbf{I_{2^{N-2}}} \\+ \mathbf{J_{z,N}^{2}}} & \mathbf{J_{N}^{2}}\\
        \mathbf{J_{xy,N}^{2}}^{\top} & \substack{\mathbf{{H}_{N-1}^{D1}} + B_{N}^{z}\mathbf{I_{2^{N-2}}}  \\+ \mathbf{J_{z,N}^{2}}}
    \end{matrix}
\end{array} \right]
\label{HIT-recursive}
\end{align}
\end{widetext}
 where  $\mathbf{I_{2^{N}}}$ denotes an identity matrix of size $2^{N}$. The quantities $\mathbf{J_{xy,N}^{1}}$ and $\mathbf{J_{xy,N}^{2}}$ are matrices that appear in the recursive definition of the diagonal blocks, labeled with superscripts $\bf{D_{1}}$ and $\bf{D_{2}}$ respectively, and contain intersite coupling of the $N^{th}$ spin site with the remaining $N-1$ sites.
To arrive at the matrix elements belonging to  $\bra{\tilde{\lambda}}\mathbf{J_{xy,N}^1}\ket{\tilde{\lambda}^\prime}$ and $\bra{\tilde{\lambda}}\mathbf{J_{xy,N}^2}\ket{\tilde{\lambda}^\prime}$ in the equation above, 
a bitwise XOR operation is constructed between the corresponding computational bases, $\ket{\tilde{\lambda}}$  and $\ket{\tilde{\lambda}^\prime}$. The XOR operation provides the identity of the spin sites where the computational basis vectors $\ket{\tilde{\lambda}}$  and $\ket{\tilde{\lambda}^\prime}$ differ, that is when the spin states are flipped between $\ket{\tilde{\lambda}}$  and $\ket{\tilde{\lambda}^\prime}$.
When the bases differ at two spin lattice site locations, $i$ and $j$, the corresponding matrix element of $\mathbf{J_{xy,N}^1}$ or $\mathbf{J_{xy,N}^2}$ 
is given by $J_{ij}^{x}\pm J_{ij}^{y}$.
The phase preceding the  $J_{ij}^{y}$ values results from an XNOR operation on the $i, j$ lattice sites discovered through the XOR operation above. 

The terms, ${\mathbf{J_{z,N}^{1}}}$ and ${\mathbf{J_{z,N}^{2}}}$, in Eq. (\ref{HIT-recursive}) are also defined in a similar fashion. Both ${\mathbf{J_{z,N}^{1}}}$ and ${\mathbf{J_{z,N}^{2}}}$ matrices are diagonal in form. Thus, the diagonal elements of $\mathbf{H_{N}^{D1}}$ and $\mathbf{H_{N}^{D2}}$ are incremented by a linear combination of all possible intersite couplings of the $N^{th}$ spin site with the remaining $N-1$ sites given by, $\sum\limits_{i=1}^{N-1}(-1)^{\tilde{\lambda}_{i}\oplus \tilde{\lambda}_{N}}J_{iN}^{z}$. 

For symmetric potentials, as noted in the paper, setting all the  transverse local qubit magnetic fields, $B_i^x$ and $B_i^y$ values to zero in Eq. (\ref{HIT-gen}), block-diagonalizes Eq. (\ref{HIT-recursive}), and 
may be recursively written as,
\begin{widetext}
\begin{align}
\mathbf{H_{N}} = 
\begin{bmatrix}
\substack{\mathbf{H_{N-1}^{D1}} + B_{N}^{z}\mathbf{I_{2^{N-2}}}\\+\mathbf{J_{z,N}^{1}}} & \mathbf{J_{xy,N}^{1}
} & \mathbf{0} & \mathbf{0}\\
\mathbf{J_{xy,N}^{1}
}^\top & \substack{\mathbf{H_{N-1}^{D2}} - B_{N}^{z}\mathbf{I_{2^{N-2}}}\\+\mathbf{J_{z,N}^{1}}} & \mathbf{0} & \mathbf{0}\\
\mathbf{0} & \mathbf{0} & \substack{\mathbf{H_{N-1}^{D2}} + B_{N}^{z}\mathbf{I_{2^{N-2}}}\\+\mathbf{J_{z,N}^{2}}} & \mathbf{J_{xy,N}^{2}
}\\
\mathbf{0} & \mathbf{0} & \mathbf{J_{xy,N}^{2}
}^\top & \substack{\mathbf{H_{N-1}^{D1}} - B_{N}^{z}\mathbf{I_{2^{N-2}}}\\+\mathbf{J_{z,N}^{2}}}\\
\end{bmatrix}
\label{HIT-nqubit-Block-1}
    \end{align}
\end{widetext}

For $N$ qubits, the number of ion-trap handles in Eq. (\ref{HIT-gen}) 
that control various sectors of the Hamiltonian matrix scale as 
\begin{align}
    \left\{ N(N+1)/2 \right\} + \left\{ N(N-1) \right\}  \rightarrow {\cal O} \left( N^2 \right),
    \label{IT-handles1}
\end{align}
Here (a) the first quantity, $\left\{ N(N+1)/2 \right\}$, refers to the parameters, $\left\{ B_{i}^z;  J_{ij}^z \right\}$, that control the diagonal elements of the matrix, and (b) the second quantity on the left, $\left\{ N(N-1) \right\}$, refers to the parameters, $\left\{ J_{ij}^x \pm J_{ij}^y \right\}$, that control the coupling between the basis vectors inside each block. 
This characterization not only elucidates the degrees of freedom of the Ising model Hamiltonian in Eq. (\ref{HIT-gen}),
but also provides the sectored availability of these control parameters. 

\bigskip

\noindent\textbf{Numerical Hamiltonians for DMANH$^+$} --
For the DMANH$^+$ system used in this work, discretized to 8 grid points, the numerical Hamiltonian reads (in units of milliHartrees):
\begin{widetext}
\begin{equation*}
    {H = \begin{pmatrix}
     37.72 & -7.478 &  0.5691 &  0.2715 & -0.2739 &  0.1491 & -0.0584 &  0.0168\\
 -7.478 &  7.050 & -7.478 &  0.5691 & 0.2715 & -0.2739 &  0.1491 & -0.0584\\
  0.5691 & -7.478 &  0.00197 & -7.478 &  0.5691 & 0.2715 & -0.2739 &  0.1491\\
0.2715 &  0.5691 & -7.478 &  0.0168 & -7.478 &  0.5691 & 0.2715 & -0.2739\\
 -0.2739 &  0.2715 &  0.5691 & -7.478 &  0.0190 & -7.478 &  0.5691 & 0.2715\\
  0.1491 & -0.2739 &  0.2715 &  0.5691 & -7.478 &  0 & -7.478 &  0.5691\\
 -0.0584 &   0.1491 & -0.2739 &  0.2715 &  0.5691 & -7.478 &  7.015 & -7.478\\
  0.0168 & -0.0584 &   0.1491 & -0.2739 &  0.2715 &  0.5691 & -7.478 &  37.60\\
    \end{pmatrix}}
\end{equation*}
\end{widetext}

Following the mapping to block-diagonal form using Eqs. 3 and 4 of the main text, the transformed Hamiltonian reads (in units of milliHartrees):
\begin{widetext}
\begin{equation*}
    {\Tilde{H} = \begin{pmatrix}
     37.68 & -7.537 &  0.7182 & -0.0024 &  0 &  0 &  0 & 0 \\
 -7.5367 &  7.182 & -7.752 &  0.8407 &  0 &  0 &  0 & 0\\
  0.7182 & -7.752 &  0.2725 & -6.909 &  0 &  0 &  0 & 0\\
 -0.0024 &  0.8407 & -6.909 & -7.460 &  0 &  0 &  0 & 0\\
 0 &  0 &  0 & 0 &  7.4960 & -8.0473 &  0.2976 &  0.5454\\
0 &  0 &  0 & 0 & -8.0473 & -0.2705 & -7.204 &  0.4200\\
0 &  0 &  0 & 0 &  0.2976 & -7.204 &  6.884 & -7.420\\
0 &  0 &  0 & 0 &  0.5454 &  0.4200 & -7.420 &  37.64\\
    \end{pmatrix}}
\end{equation*}
\end{widetext}

